\documentclass[onecolumn, useAMS]{mn2e}
\usepackage{graphicx}
\usepackage{amsfonts}
\usepackage{subfigure}
\usepackage{color}
\usepackage{times}

\usepackage{ifpdf}
\ifpdf
\usepackage[hyperfootnotes=false,colorlinks=true]{hyperref}
\hypersetup{linkcolor=red,citecolor=blue}
\usepackage{epstopdf}
\fi

\newcommand{\B}{\beta}
\newcommand{\delT}{\Delta T}
\newcommand{\delTobs}{\widetilde{\Delta T}}

\newcommand{\lp}{{l^\prime}}
\newcommand{\Lp}{{L^\prime}}
\newcommand{\mpp}{m^{\prime\prime}}
\newcommand{\lpp}{{l^{\prime\prime}}}
\newcommand{\Mpp}{{M^{\prime\prime}}}
\newcommand{\p}{\prime}
\newcommand{\q}{\mathbf{\hat{q}}}
\newcommand{\qp}{\mathbf{\hat{q}^\prime}}

\newcommand{\cclobs}{\langle \widetilde{\mathcal{C}}_l \rangle}

\newcommand{\intq}{\int_{4\pi} d\Omega_{\hat{\mathbf{q}}}}

\newcommand{\intqa}{\int_{4\pi} d\Omega_{\hat{\mathbf{q}}_1}}
\newcommand{\intqb}{\int_{4\pi} d\Omega_{\hat{\mathbf{q}}_2}}

\newcommand{\mat}[2][cccccccccccccccccccccccccccccccccccc]{\left(
\begin{array}{#1}
#2\\
\end{array}
\right)}
\newcommand{\wjjj}[6]
{{ \left(
\begin{array}{lcr} #1 & #2 & #3 \\#4 & #5 & #6 \end{array}
\right) }}

\def\lsim{\mathrel{\rlap{\lower4pt\hbox{\hskip1pt$\sim$}}
    \raise1pt\hbox{$<$}}}                
\def\gsim{\mathrel{\rlap{\lower4pt\hbox{\hskip1pt$\sim$}}
    \raise1pt\hbox{$>$}}}                

\begin{document}
\title[CMB power spectrum estimation with ...]
{CMB power spectrum estimation with non-circular beam and
incomplete sky coverage}

\author[Mitra, Sengupta, Ray, Saha \& Souradeep]
{Sanjit Mitra$^{1,2}$ Anand S. Sengupta$^{3}$ Subharthi
Ray$^{1,4}$ Rajib Saha$^{1,5,2}$
\newauthor Tarun Souradeep$^{1}$ \\
$^1$ Inter-University Centre for Astronomy and Astrophysics,
Post Bag 4, Ganeshkhind, Pune 411007, India.\\
$^2$ Jet Propulsion Laboratory, California Institute of
Technology, Pasadena, CA 91109, USA \\
$^3$ LIGO Laboratory, California Institute of Technology, Pasadena CA 91125, USA.\\
$^4$ Astrophysics and Cosmology Research Unit, School of
Mathematical Sciences, University of KwaZulu-Natal, \\
Private Bag X54001, Durban 4000, South Africa\\
$^5$ Indian Institute of Technology, Kanpur, Kanpur 208 016,
India }

\maketitle

\begin{abstract}

Over the last decade, measurements of the CMB anisotropy has
spearheaded the remarkable transition of cosmology into a
precision science. However, addressing the systematic effects in
the increasingly sensitive, high resolution, `full' sky
measurements from different CMB experiments pose a stiff
challenge. The analysis techniques must not only be
computationally fast to contend with the huge size of the data,
but, the higher sensitivity also limits the simplifying
assumptions which can then be invoked to achieve the desired speed
without compromising the final precision goals. While maximum
likelihood is desirable, the enormous computational cost makes
the suboptimal method of power spectrum estimation using
Pseudo-C$_l$ unavoidable for high resolution data. The debiasing
of the Pseudo-C$_l$ needs account for non-circular beams, together
with non-uniform sky coverage. We provide a (semi)analytic
framework to estimate bias in the power spectrum due to the
effect of beam non-circularity and non-uniform sky coverage
including incomplete/masked sky maps and scan strategy. The
approach is perturbative in the distortion of the beam from
non-circularity, allowing for rapid computations when the beam is
mildly non-circular. We advocate that it is computationally
advantageous to employ `soft' azimuthally apodized masks whose
spherical harmonic transform die down fast with $m$. We
numerically implement our method for {\em non-rotating beams}. We
present preliminary estimates of the computational cost to
evaluate the bias for the upcoming CMB anisotropy probes
($l_{\rm{max}} \sim 3000$), with angular resolution comparable to
the Planck surveyor mission. We further show that this
implementation and estimate is applicable for rotating beams on
equal declination scans and possibly can be extended to simple
approximations to other scan strategies.

\end{abstract}


\section{Introduction}

The fluctuations in the Cosmic Microwave Background (CMB)
radiation are theoretically very well understood, allowing
precise and unambiguous predictions for a given cosmological
model (Bond 1996, Hu and Dodelson 2002). The measurement of CMB
anisotropy with the ongoing Wilkinson Microwave Anisotropy Probe
(WMAP) and the upcoming Planck surveyor, has ushered in a new era
of precision cosmology. Such data rich experiments, with
increased sensitivity, high resolution and `full' sky
measurements pose a stiff challenge for current analysis
techniques to realize the full potential of precise determination
of cosmological parameters. The analysis techniques must not only
be computationally fast to contend with the huge size of the
data, but, the higher sensitivity also limits the simplifying
assumptions that can then be invoked to achieve the desired speed
without compromising the final accuracy. As experiments improve
in sensitivity, the inadequacy in modeling the observational
reality start to limit the returns from these experiments. The
current effort is to push the boundary of this inherent
compromise faced by the current CMB experiments that measure the
anisotropy in the CMB temperature and polarization.

Accurate estimation of the angular power spectrum, $C_{l}$, is
inarguably the foremost concern of most CMB experiments. The
extensive literature on this topic has been summarized (Hu and
Dodelson 2002, Bond 1996, Efstathiou 2004). For Gaussian,
statistically isotropic CMB sky, the $C_l$ that corresponds to
the covariance that maximizes the multivariate Gaussian PDF of
the temperature map, $\Delta T(\q)$, is the Maximum Likelihood
(ML) solution. Different ML estimators have been proposed and
implemented on CMB data of small and modest sizes (Gorski 1994,
Gorski et al. 1994, 1996, 1997, Tegmark 1997, Bond et al. 1998).
While it is desirable to use optimal estimators of $C_l$ that
obtain (or iterate toward) the ML solution for the given data,
these methods are usually limited by the computational expense of
matrix inversion that scales as $N_d^3$ with data size $N_d$
(Borrill 1999, Bond et al. 1999). Various strategies for speeding
up ML estimation have been proposed, such as, exploiting the
symmetries of the scan strategy (Oh et al. 1999), using
hierarchical decomposition (Dore et al. 2001), iterative
multi-grid method (Pen 2003), etc. Variants employing linear
combinations of $\Delta T(\q)$ such as $a_{lm}$ on set of rings
in the sky can alleviate the computational demands in special
cases (Challinor et al. 2002, van Leeuwen et al. 2002, Wandelt \&
Hansen 2003). Other promising `exact' power estimation methods
have been recently proposed(Knox et al. 2001, Jewell et al. 2004,
Wandelt 2004).

However there also exist computationally rapid, sub-optimal
estimators of $C_l$. Exploiting the fast spherical harmonic
transform ($\sim N_d^{3/2}$), it is possible to estimate the
angular power spectrum $C_l= \sum_m |a_{lm}|^2/(2l+1)$ rapidly
(Yu \& Peebles 1969, Peebles 1974). This is commonly referred to
as the Pseudo-$C_l$ method (Wandelt et al. 2003). Analogous
approach employing fast estimation of the correlation function
$C(\q\cdot\qp)\equiv \langle \Delta T(q) \Delta T(q')\rangle$
have also been explored (Szapudi et al. 2001, Szapudi et al.
2001). It has been recently argued that the need for optimal
estimators may have been over-emphasized since they are
computationally prohibitive at large $l$. Sub-optimal estimators
are computationally tractable and tend to be nearly optimal in
the relevant high $l$ regime. Moreover, already the data size of
the current sensitive, high resolution, `full sky' CMB
experiments, such as WMAP have been compelled to use sub-optimal
Pseudo-$C_l$ and related methods (Bennett et al. 2003, Hinshaw et
al. 2003). On the other hand, optimal ML estimators can readily
incorporate and account for various systematic effects, such as
noise correlations, non-uniform sky coverage and beam
asymmetries. A hybrid approach of using ML estimation of $C_l$
for low $l$ for low resolution map and Pseudo-$C_l$ like
estimation for large $l$ where it is nearly optimal has been
suggested (Efstathiou 2004) and has even been employed by the
recent analysis of the WMAP-3 year data (Hinshaw et al. 2007).
The systematic correction to the Pseudo-$C_l$ power spectrum
estimate arising from non-uniform sky coverage has been studied
and implemented for CMB temperature (Hivon et al. 2002) and
polarization (Brown at al 2005). The {\it leading} order
systematic bias due to noncircular beam has been studied by us in
an earlier publication (Mitra et al. 2004). Here we extend the
results in a thorough analytical approach to include all the
significant perturbation orders and combine the effect of
non-uniform sky coverage.

It has been usual in CMB data analysis to assume the experimental beam
response to be circularly symmetric around the pointing
direction. However, real beam response functions have deviations
from circular symmetry. Even the main lobe of the beam response of
experiments are generically non-circular (non-axisymmetric) since
detectors have to be placed off-axis on the focal plane. (Side lobes
and stray light contamination add to the breakdown of this
assumption). For highly sensitive experiments, the systematic errors
arising from the beam non-circularity become progressively more
important.  Dropping the circular beam assumption leads to  major
complications at every stage of the analysis pipeline. The extent to which
the non-circularity of the beam affects the step of going from the time-stream
data to sky map is very sensitive to the scan-strategy. The beam now
has an orientation with respect to the scan path that can potentially
vary along the path. This introduces an additional complication - that the
beam function is inherently time dependent and difficult to deconvolve.

\begin{figure*}
\centering
\includegraphics[angle = 0,scale = 0.37 ]{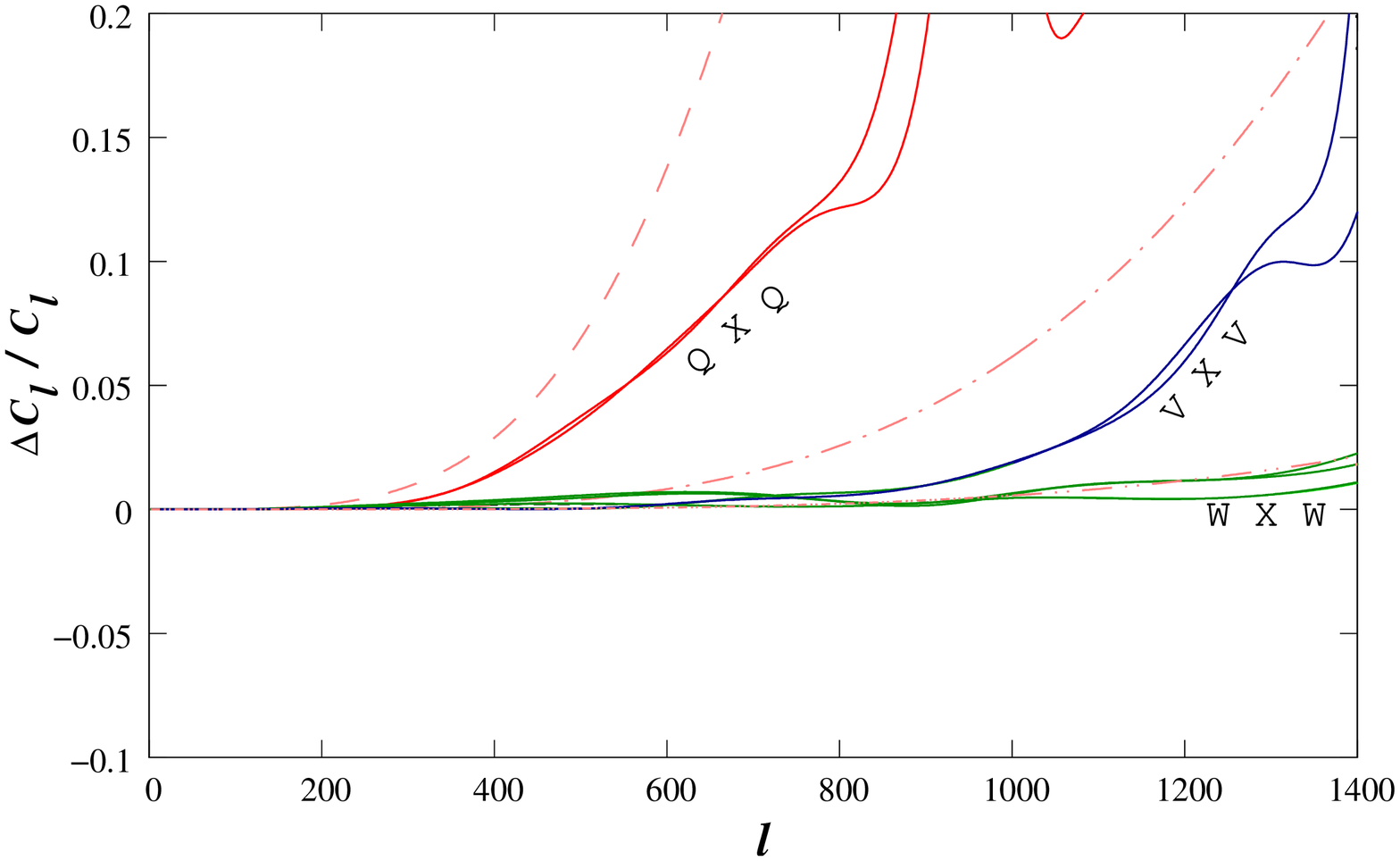}
\includegraphics[angle = 0,scale = 0.35 ]{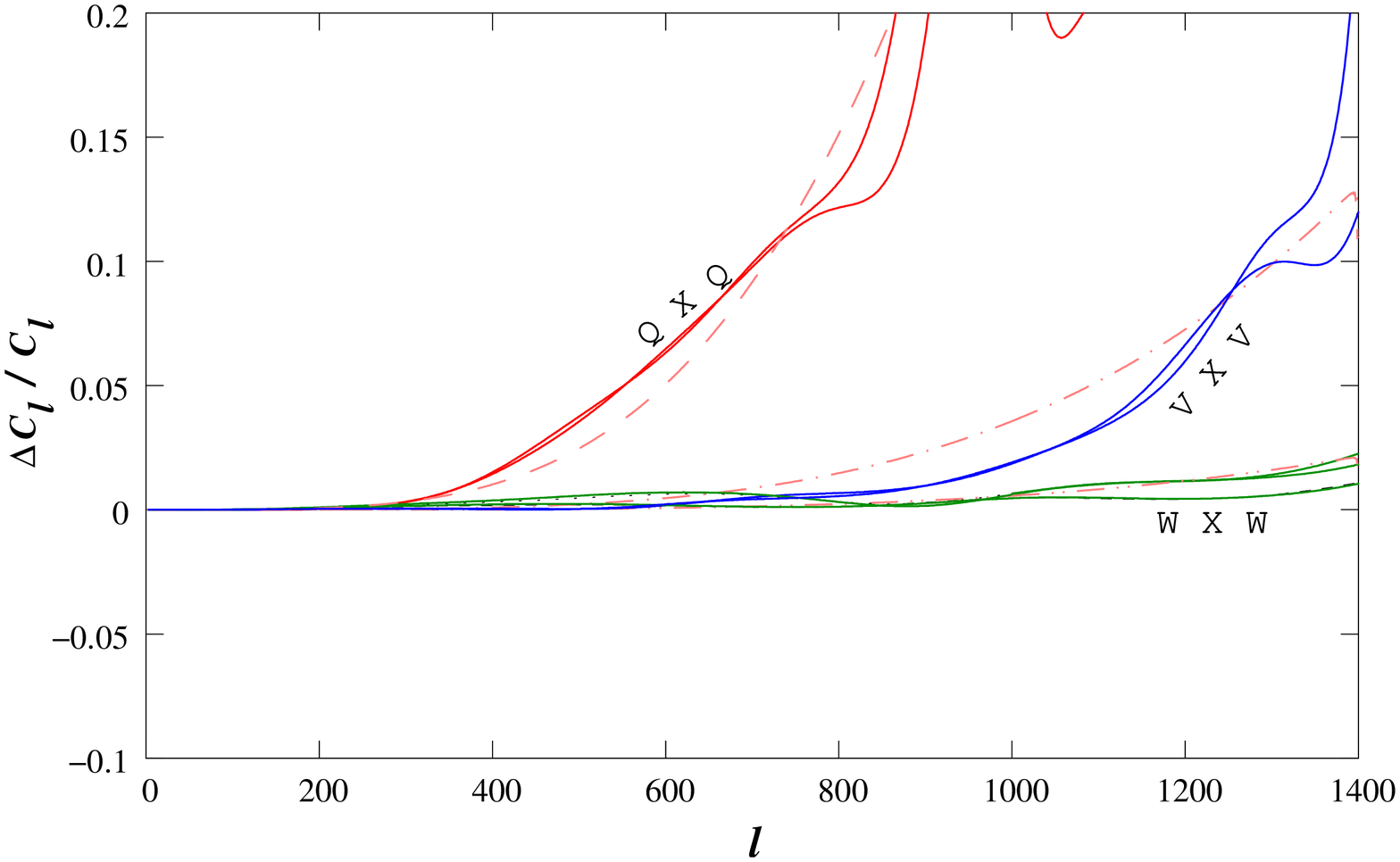}
\caption{The predicted non-circular beam correction for a CMB
experiment with elliptical Gaussian beam with {\it fwhm}
beam-width of $0.51^\circ$ and eccentricity $e := \sqrt{1 -
b^2/a^2}$, where $a$ ($b$) is the semi-major (semi-minor) axis,
$e=0.65$ for the  Q beam (dashed line), beam-width of
$0.35^\circ$ and $e=0.46$ for the V beam (dash-dotted line) and
beam-width of $0.22^\circ$  and $e=0.4$  for the W beam
(dash-dot-dot line) are shown in the left plot. The solid curves
are the non circular beam corrections estimated by the WMAP team
for the Q,V and W channels. In the plot in the right, we show
that the non-circular beam correction matches with the estimates
of the WMAP team for a slightly different change in the
eccentricity of the Q and V beams (W beams matched in the left
plot), with the new eccentricities $e=0.50$ and $e=0.40$
respectively. This difference is attributed to the fact that the
beams visit the pixels multiple times with different
orientations, and hence the {\it effective} eccentricity is
reduced. }

\label{clerr}
\end{figure*}

Even after a sky map is made, the non-circularity of the effective
beam affects the estimation of the angular power spectrum, $C_l$,
by coupling the power at different multipoles, typically, on
scales beyond the inverse angular beam-width.  Mild deviations
from circularity can be addressed by a perturbation approach
(Souradeep \& Ratra 2001, Fosalba et al. 2002) and the effect of
non-circularity on the estimation of CMB power spectrum can be
studied (semi) analytically (Mitra et al. 2004).
Figure~\ref{clerr} shows the predicted level of bias due to
noncircular beams in our formalism for elliptical beams
compared to the noncircular beam corrections computed in the
recent data release by WMAP (Hinshaw et al. 2007).

To avoid contamination of the primordial CMB signal by Galactic
emission, the region around the Galactic plane is masked from
maps. If the Galactic cut is small enough, then the coupling
matrix will be invertible, and the two-point correlation function
can be determined on all angular scales from the data within the
uncut sky (Mortlock et al. 2002).  Hivon et al. (2002) present a
technique (MASTER) for fast computation of the power spectrum
accounting for the galactic cut, but restricted to circular
beams. In our present work, we present analytical expressions for
the bias matrix of the Pseudo-${C_l}$ estimator for the
incomplete sky coverage, \emph{using a noncircular beam}. In
Section~\ref{form} we show a heuristic approach to the analytic
form of the bias matrix taking into account the above mentioned
effects. We have shown that our estimation matches with the
existing results in different limits in Section~\ref{sec:limit}.
In Section~\ref{num} we outline the numerical implementation of
our approach, estimate the computational cost and suggest a
potential algorithmic route to reducing the cost. The discussion
and conclusion of this work is given in Section~\ref{disc}.

\section{Formalism}\label{form}

The CMB temperature anisotropy field $\delT(\q)$
over all the sky directions $\q \equiv (\theta,\phi)$ is assumed to be Gaussian
and statistically isotropic and hence the angular power spectrum stores all the
statistical information about the anisotropy field. These temperature fluctuations are
expanded in the basis of spherical harmonics,

\begin{equation}
\delT(\q) \ = \ \sum_{lm} a_{lm} \, Y_{lm}(\q),
\end{equation}
where $a_{lm}$  are the spherical harmonic transforms of the temperature anisotropy field
\begin{equation}
a_{lm} \ := \ \intq \, \delT(\q) \, Y_{lm}^*(\q).
\end{equation}

The observed CMB temperature fluctuation field $\delTobs(\q)$ is a
convolution of the ``beam" profile $B(\q,\qp)$ with the real
temperature fluctuation field $\delT(\q^\prime)$ and contaminated by additive experimental
noise $n(\q)$. Moreover, due to the strong influence of the foreground emission in our galactic
plane and point sources, some of the pixels have to be masked prior to power spectrum
estimation. The mask function $U(\q)$ is usually assigned zero for the corrupt
pixels and one for the clean ones, but it could be a smoother weight function also,
that can take values between zero and one, as long as it sufficiently masks the
foreground contamination. Mathematically the observed temperature can be expressed as
\begin{equation}
\delTobs(\q) \ = \ U(\q) \left[ \int_{4\pi} d\Omega_{\qp} \, B(\q,\qp) \,
\delT(\qp) \ + \ n(\q) \right].
\end{equation}
The spherical harmonic transform of the mask function,
\begin{equation}
U_{lm}  \ \equiv \ \int_{4\pi} d\Omega_{\mathbf{\hat{q}}} \,
Y_{lm}^*({\mathbf{\hat{q}}}) \, U(\mathbf{\hat{q}})\,,
\end{equation}
is a very useful quantity for our analysis.

Statistical isotropy of underlying CMB anisotropy signal implies that the two point correlation function
$C(\q_1,\q_2) = \langle \delT(\q_1) \delT(\q_2) \rangle$ depends only
on the angular separation of the direction, i.e.,
$C(\q_1,\q_2) = C(\q_1 \cdot \q_2)$. We can therefore expand it
as a Fourier-Legendre series
\begin{equation}
\langle \delT(\q_1) \delT(\q_2) \rangle \ = \ \sum_{l=0}^\infty
\frac{2l+1}{4\pi} C_l \, P_l(\q_1 \cdot \q_2),
\end{equation}
where the coefficients $C_l$ constitute the CMB power spectrum
\begin{equation}
C_l \ \equiv \ \intqa \intqb  \langle \delT(\q_1) \, \delT(\q_2) \rangle \, P_l(\q_1 \cdot \q_2).
\end{equation}
The addition theorem for spherical harmonics
\begin{equation}
\frac{4\pi}{2l+1} \sum_{m=-l}^l Y_{lm}^*(\q_1) \, Y_{lm}(\q_2) \ = \ P_l(\q_1\cdot\q_2),
\end{equation}
and the orthogonality of Legendre polynomials
\begin{equation}
\int_{-1}^{1} dx \, P_l(x) \, P_{l'}(x) \ = \ \frac{2}{2l+1} \, \delta_{ll'}
\end{equation}
can then be used to show that the matrix $\langle a_{lm} a_{l'm'} \rangle $ is diagonal:
\begin{equation}
\langle a_{lm} a_{l'm'} \rangle \ = \ C_l \, \delta_{ll'} \, \delta_{mm'}.
\end{equation}

The observed two point correlation function for a statistically isotropic CMB
anisotropy signal is
\begin{equation}
\widetilde{C}(\q,\q^\prime) =  \langle \delTobs(\q) \delTobs(\qp)
\rangle =   \sum^\infty_{l = 0} {(2 l + 1) \over 4\pi}\, C_l\,\,
   W_l(\q,\,\q^\prime )\,,
   \label{corre}
\end{equation}
where $C_l$ is the angular power spectrum of CMB anisotropy signal and
the \emph{window} function,
\begin{equation}
W_l(\q_1,\q_2) \ := \ \int d\Omega_\q \int d\Omega_\qp \,
B(\q_1, \q) B(\q_2, \qp) P_l(\q \cdot \qp), \label{eq:defW}
\label{windef}
\end{equation}
encodes the effect of finite resolution through the beam function.
A CMB anisotropy experiment probes a range of angular scales
characterized by this window function $W_l(\q,\q^\prime)$.  The
window depends both on the scanning strategy as well as the
angular resolution and response of the experiment. However, it is
neater to logically separate these two effects by expressing the
window $W_l(\q,\q^\prime)$ as a sum of `elementary' window
functions of the CMB anisotropy (Souradeep \& Ratra 2001). For a
given scanning strategy, the results can be readily generalized
using the representation of the window function as a sum over
elementary window functions (see, {\it e.g.,} Souradeep \& Ratra
2001, Coble et al. 2003, Mukherjee et al. 2003).

For some experiments, the beam function may be assumed to be
circularly symmetric about the pointing direction, i.e., $B(\q,
\q^\prime) \equiv B(\q\cdot\q^\prime)$, without significantly
affecting the results of the analysis.  In any case, this
assumption allows a great simplification, since the beam function
can then be represented by an expansion in Legendre polynomials as
\begin{equation}
B(\q\cdot\qp) \ = \ \frac{1}{4\pi}\,\sum_{l=0}^\infty\, (2l+1)\,
B_l\, P_l(\q\cdot\qp), \label{eq:BthetafromBl}
\end{equation}
where,
\begin{equation}
B_l  \equiv \int_{-1}^{1} d(\q \cdot \qp) \, P_l(\q \cdot\qp)\, \mathcal{B}(\q \cdot \qp),
\label{eq:bldef}
\end{equation}
and $\mathcal{B}(\q \cdot \qp)$ is the \textit{circularized}
beam obtained by averaging $B(\mathbf{\hat{z}},\mathbf{\hat{q}})$
over the azimuthal coordinate $\phi$.
Consequently, it is straightforward to derive the well known
simple expression
\begin{equation}
   W_l(\q,\,\qp) \ = \ B^2_l \, P_l(\q\cdot\qp)\,,
   \label{isowine}
\end{equation}
for a circularly symmetric beam function. For non-circular beams
and incomplete sky coverage the Pseudo-$C_l$ estimator,
\begin{equation}
\widetilde{{C}}_l \ \equiv \ \frac{1}{2l+1}\sum_{m=-l}^l \left|\widetilde{a}_{lm}\right|^2,
\end{equation}
takes the form
\begin{equation}
\widetilde{{C}}_l \ = \ \frac{1}{4\pi}
\int_{4\pi} d\Omega_{\q_1} \int_{4\pi} d\Omega_{\q_2} \, U(\q_1)
\, U(\q_2) \, P_l(\q_1 \cdot \q_2) \, \delTobs(\q_1) \,
\delTobs(\q_2).
\end{equation}
In this case, the expectation value of the Pseudo-$C_l$ estimator becomes
\begin{equation}
\cclobs \ = \ \sum_{\lp} A_{l\lp} C_\lp \ + \ \bar{C}_l^N,
\end{equation}
that is, the estimator is non-trivially biased, where the {\em
bias matrix} $A_{l\lp}$ takes the form
\begin{eqnarray}
A_{l\lp} = \frac{1}{2l+1} \sum_{n=-l}^l\!
\sum_{m=-\lp}^\lp \!\left|\int_{4\pi}\!\!\! d\Omega_{\q} U(\q) \, Y_{l
n}(\q) \left[ \int_{4\pi}\!\!\! d\Omega_{\qp} \, Y^*_{\lp m}(\qp) \,
B(\q,\qp) \right] \right|^2.
\label{eq:pscl1}
\end{eqnarray}
The noise term $\bar{C}_l^N$, arising from instrumental noise,
can be measured to a very high accuracy. If the noise term for
full sky coverage $C_l^N$ is known, it can be combined with the
bias matrix for incomplete sky coverage $M_{l\lp}$ (Hivon et al.
2002) to obtain the noise power spectrum for cut-sky $\bar{C}_l^N$
\begin{equation}
\bar{C}_l^N \ = \ \sum_{\lp} M_{l\lp} \, C_\lp^N,
\end{equation}
where, $M_{l\lp}$ is defined as
\begin{equation}
M_{l\lp} \ =  \, \frac{2l' +1}{4\pi} \sum_{l''=|l-l'|}^{l+l'} \, (2l''+1)
\, \wjjj{l}{l'}{l''}{0}{0}{0}^2 \, {\cal{U}}_{l''},
\end{equation}
with ${\cal{U}}_{\lpp} \equiv \sum_{\mpp=-\lpp}^{\lpp} |U_{\lpp\mpp}|^2/(2\lpp+1)$.
Computation of the bias matrix is important for defining the
unbiased Pseudo-$C_l$ estimator
\begin{equation}
\widetilde{C}_l^{\rm{UB}} \ \equiv \ \sum_{\lp} [A^{-1}]_{l\lp} \,
\left( \widetilde{C}_\lp \ - \ \sum_{\lpp} M_{\lp\lpp} \,
C_\lpp^N \right)
\end{equation}
that removes the systematic effects of beam non-circularity and
incomplete sky coverage.

The experimental beams in CMB experiments are mildly
non-circular, and hence it makes sense to define Beam Distortion
Parameters (BDP) $\beta_{lm}(\ll 1)$ so that we can calculate the
result in a perturbative expansion of small parameter. The BDP
$\beta_{lm}$ is expressed as $\beta_{lm} \equiv b_{lm}/b_{l0}$,
where the $b_{lm}$ are the spherical harmonic moments of the beam
function for the pointing direction $\mathbf{\hat{z}}$:
\begin{equation}
b_{lm}  \equiv \int_{4\pi} d\Omega_{\mathbf{\hat{q}}} \,
Y_{lm}^*({\mathbf{\hat{q}}}) \, B(\mathbf{\hat{z}},\mathbf{\hat{q}})\, , \label{bdp}
\end{equation}
and $B_l$, defined in Eq.~(\ref{eq:bldef}), is connected to
$b_{l0}$ through the relation:
\begin{equation}
\nonumber
B_l  = \int_{0}^{\pi} \sin\theta d\theta \,
\sqrt{\frac{4\pi}{2l+1}} \, Y_{l0}^*({\mathbf{\hat{q}}}) \left[
\frac{1}{2\pi} \int_{0}^{2\pi} d\phi \, B(\mathbf{\hat{z}},\mathbf{\hat{q}}) \right]
= \sqrt{\frac{4\pi}{2l+1}} \, b_{l0}. \label{Blbl0}
\end{equation}
Evaluation of the spherical harmonic transforms of the beam
function $B(\q,\qp)$ for each pointing direction $\q$ is
computationally prohibitive. We use the spherical harmonic
transforms  $b_{lm}$ of $B(\hat{\mathbf{z}},\qp)$, incorporating
rotation in it via Wigner-$D$ functions, in order to compute the
harmonic transforms of $B(\q,\qp)$ for any $\q$ by using the
formula \cite{TR:2001}

\begin{equation}
\int_{4\pi} d\Omega_{\qp} \, Y^*_{\lp m}(\qp) \, B(\q,\qp)  \ = \
\sqrt{\frac{2\lp+1}{4\pi}} \sum_{m^\p=-\lp}^\lp B_\lp \,
\beta_{\lp m^\p} \, D^\lp_{mm^\p}(\q,\rho(\q)).
\end{equation}
Then, using the spherical harmonic expansion of the
mask function $U(\q)$
\begin{equation}
U(\q) \ = \ \sum_{l=0}^{\infty} \sum_{m=-l}^{l} U_{lm} \,
Y_{lm}(\q), \label{maskTr}
\end{equation}
we can rewrite the general form of the bias matrix
[Eq.~(\ref{eq:pscl1})] as
\begin{equation}
A_{l l^\prime} \ = \ \frac{B_l^2}{4\pi}
\frac{(2l^\prime+1)}{(2l+1)}
\sum_{n=-l}^{l} \sum_{m=-l^\prime}^{l^\prime}
\left|\sum_{m^\prime=-l^\prime}^{l^\prime} \beta_{l^\prime
m^\prime} \sum_{\lpp =0}^{\infty} \sum_{\mpp =-\lpp}^{\lpp}
U_{\lpp\mpp} \, J^{l\lpp\lp}_{n\mpp mm^\prime} \right|^2,
\label{eq:gen}
\end{equation}
where
\begin{eqnarray}
J^{l\lpp\lp}_{n\mpp mm^\prime} &:=& \int_{4\pi}
d\Omega_{\mathbf{\hat{q}}} \, Y_{ln}(\mathbf{\hat{q}}) \,
Y_{\lpp\mpp}(\mathbf{\hat{q}}) \,
D_{mm^\prime}^{l^\prime}(\mathbf{\hat{q}},\rho(\mathbf{\hat{q}})) \label{eq:defJ}\\
&=&(-1)^{n+\mpp}
\frac{\sqrt{(2l+1)(2\lpp+1)}}{4\pi}\sum_{L=|l-\lpp|}^{l+\lpp}
C^{L0}_{l0\lpp 0} \, C^{L(n+\mpp)}_{ln\lpp \mpp}
\times \nonumber\\ &&  \sum_{\Lp=|L-\lp|}^{L+\lp}
C^{\Lp(m-n-\mpp)}_{L(-n-\mpp)\lp m} \, C^{\Lp m^\p}_{L0\lp m^\p} \, \chi^{L'}_{(m-n-\mpp)m'}[\rho(\q)]
\label{eq:JClebsch}
\end{eqnarray}
and
\begin{equation}
\chi^{l}_{mm'}[\rho(\q)] \ := \ \int_{4\pi} d\Omega_{\mathbf{\hat{q}}} \,
D^{l}_{mm'}(\mathbf{\hat{q}},\rho(\mathbf{\hat{q}})).
\end{equation}
To obtain the expression for the $J$ coefficients above we have
introduced Clebsch Gordon coefficients
$C^{ll'}_{m_{1}m_{2}m'_{1}m'_{2}}$ along with sinusoidal
expansion of Wigner-$d$ (see Appendix \ref{appendix:clebs} for
details).

Eq.~(\ref{eq:JClebsch}) provides the expression for the bias
[Eq.~(\ref{eq:gen})] in its full generality. For a specified
general form of the scan strategy, however, we need to precompute
the coefficients $\chi^l_{mm'}$ which are functionals of
$\rho(\q)$.
Special cases, which often provide good approximations to the
real scan strategies, provide significant computational
advantages. From the definition of $\chi$ coefficients we show
below that
\renewcommand{\labelenumi}{\Roman{enumi}.}
\begin{enumerate}
\item For a $\rho(\q)$ separable in declination and Right
Ascension parts, i.e. $\rho(\q) \equiv \Theta(\theta) +
\Phi(\phi)$,
\begin{equation}
\chi^{l}_{mm'}[\rho(\q)] \ = \ \int_0^{2\pi}   d\phi \,e^{-i m
\phi} \, e^{-i m' \Phi(\phi)} \int_0^\pi \sin\theta \, d\theta \,
d^l_{mm'}(\theta) \, e^{-i m' \Theta(\theta)}
\end{equation}
would have significant values only in a much constrained domain
of the indices $m,m'$.
\item For equal declination scan strategies, where $\rho(\q) \equiv
\rho(\theta)$, we show below that the computational cost reduces to that
of the non-rotating beam ($\rho(\q)=0$) case.
\end{enumerate}
Hence, the study of the bias computation for {\em non-rotating beams}
provides an analytic framework that is computationally equivalent to
a broader class of scan strategies as mentioned at the appropriate
places below.

It is clear from the definition of $\chi^l_{mm'}$ that for
equal declination scans [case~II] the $\phi$ integral
is leading to the constraint
\begin{equation}
\chi^l_{mm'}[\rho(\q)=0] \propto \int_0^{2\pi} d\phi \, e^{i m \phi}  \
= \ 2 \pi \, \delta_{m0}.
\end{equation}
So the $J$ coefficients in Eq.~(\ref{eq:JClebsch}) contribute to
bias only when $m''=m-n$, saving huge amount of computation. For
a much broader class within case~I, it is reasonable to expect
that the $\chi$ symbols in the final expression for bias will
contribute when  $m''$ is close to $m-n$.

For equal declination scan strategies, the expression for the bias matrix can be written as
\begin{eqnarray}
A_{l l^\prime} &=& B_l^2 \, \frac{(2\lp+1)}{16\pi} \sum_{n=-l}^{l}
\sum_{m=-l^\prime}^{l^\prime} \left| \sum_{\lpp =0}^{\infty}
\sqrt{2\lpp+1} \, U_{\lpp(m-n)} \right. \ \times\nonumber\\
&&\sum_{L=|l-\lpp|}^{l+\lpp}
C^{L0}_{l0\lpp 0} \, C^{Lm}_{ln\lpp(m-n)}  \sum_{\Lp=|L-\lp|}^{L+\lp}
C^{\Lp
0}_{L-m\lp m}  \sum_{N=-\Lp}^{\Lp}
d^{\Lp}_{0N}\left(\frac{\pi}{2}\right) \ \times
\nonumber\\ && \left.
\sum_{m^\prime=-l^\prime}^{l^\prime} \beta_{l^\prime m^\prime} \,
C^{\Lp m^\p}_{L0\lp m^\p} \,
d^{\Lp}_{Nm^\p}\left(\frac{\pi}{2}\right) \, \Gamma_{m'N}[\rho(\theta)] \right|^2,
\label{eq:BiasClebschEDS}
\end{eqnarray}
where the coefficients
\begin{equation}
\Gamma_{m'N}[\rho(\theta)] \ := \ i^{m^\p} \, (-1)^N \int_0^\pi
\sin\theta d\theta \, e^{iN\theta} \, e^{-i m' \rho(\theta)} \label{eq:defGamma}
\end{equation}
are functionals of the scan strategy and have to be precomputed
(analytically/numerically) for a given $\rho(\theta)$.

For an efficient computation of the $J$ coefficients using a numerical
implementation scheme described in section~\ref{num}, we derive an
alternative expression for the $J$ coefficients using the sinusoidal
expansion of Wigner-$d$ that {\em only involves} the $d^{l}_{mm'}(\pi/2)$ symbols
(see Appendix \ref{appendix:wigd} for details):
\begin{eqnarray}
J^{l\lpp\lp}_{n\mpp mm^\p} &=& \frac{\sqrt{(2l+1)(2\lpp+1)}}{4\pi} \, i^{n+m+m''}
\sum_{M=-l}^{l} d^{l}_{nM}\left(\frac{\pi}{2}\right) \,
d^{l}_{M0}\left(\frac{\pi}{2}\right) \times
\nonumber\\
&&\sum_{\Mpp=-\lpp}^{\lpp}
d^{\lpp}_{\mpp\Mpp}\left(\frac{\pi}{2}\right) \,
d^{\lpp}_{\Mpp0}\left(\frac{\pi}{2}\right) \sum_{M^\p=-\lp}^{\lp}
d^{\lp}_{mM^\p}\left(\frac{\pi}{2}\right) \, d^{\lp}_{M^\p
m^\p}\left(\frac{\pi}{2}\right) \, \Xi_{(M+M'+M'')(m''-m+n)m'} \label{eq:JWigd}
\end{eqnarray}
where the coefficients,
\begin{equation}
\Xi_{m_1 m_2 m_3} \ := \ (-1)^{m_1} \, i^{m_3} \int_0^{2\pi} d\phi \, e^{i m_2 \phi}
\int_0^\pi \sin\theta d\theta \, e^{i m_1 \theta} \, e^{-i m_3 \rho(\q)},
\end{equation}
have to be precomputed for a given scan strategy. Since the above
expression [Eq.~(\ref{eq:JWigd})] provides just an alternate form
for the most general expression for bias, the discussion on
different special cases given after Eq.~(\ref{eq:JClebsch}) also
holds here. Thus, in case~I [$\rho(\q) \equiv \Theta(\theta) +
\Phi(\phi)$] the $\Xi$ coefficients would take the form
\begin{equation}
\Xi_{m_1 m_2 m_3} \ := \ (-1)^{m_1}  \, i^{m_3} \int_0^{2\pi}
d\phi \, e^{i m_2 \phi} \,  \, e^{-i m_3 \Phi(\phi)} \int_0^\pi
\sin\theta d\theta \, e^{i m_1 \theta} \, e^{-i m_3
\Theta(\theta)},
\end{equation}
which are likely to contribute significantly for a highly
constrained set of $m_1, m_2, m_3$ values and in case~II (equal
declination scans) $\Xi_{m_1 m_2 m_3}$ contributes only for $m_2
= 0$, so the expression for the bias matrix becomes
\begin{eqnarray}
A_{l l^\prime} &=& B_l^2 \, \frac{(2\lp+1)}{16\pi}
\sum_{n=-l}^{l} \sum_{m=-l^\prime}^{l^\prime} \left| \sum_{\lpp
=0}^{\infty} \sqrt{2\lpp+1} \, U_{\lpp(m-n)} \right. \ \times \nonumber\\
&&\sum_{\Mpp=-\lpp}^{\lpp}
d^{\lpp}_{(m-n)\Mpp}\left(\frac{\pi}{2}\right) \,
d^{\lpp}_{\Mpp0}\left(\frac{\pi}{2}\right) \sum_{M=-l}^{l}
d^{l}_{nM}\left(\frac{\pi}{2}\right) \,
d^{l}_{M0}\left(\frac{\pi}{2}\right) \ \times \nonumber\\
&& \left.  \sum_{M^\p=-\lp}^{\lp}
d^{\lp}_{mM^\p}\left(\frac{\pi}{2}\right)
\sum_{m^\prime=-l^\prime}^{l^\prime} \beta_{l^\prime m^\prime} \,
d^{\lp}_{M^\p m^\p}\left(\frac{\pi}{2}\right) \, \Gamma_{m'(M+M'+M'')}[\rho(\theta)]
\right|^2.\label{eq:BiasWigdEDS}
\end{eqnarray}
Here we have use the same $\Gamma$ coefficients as defined in Eq.~(\ref{eq:defGamma}).
In most of the cases, the beam pattern and the scan strategies
have trivial symmetries, where only the real parts of the
$\Gamma$ coefficients contribute.

The case of non-rotating beams, $\rho(\q)=0$, has been explicitly
worked out in Appendix~\ref{app:noRot}. This is a specific
example where the real part of the $\Gamma$ symbols can be
written in a closed form and the result is:
\begin{equation}
\Re\left[\Gamma_{m'N}[\rho(\q)=0]\right] \ =: \ f_{m'N} \ = \
\left\{
\begin{array}{cl} (-1)^{(m^\p\pm 1)/2} \, \pi/2        & \mbox{if
$m^\p=odd$ and $N=\pm 1$}\\ (-1)^{m^\p/2}
\, 2/(1-N^2) & \mbox{if both $m^\p,N=0$ or $even$}\\
0               & \mbox{otherwise.} \end{array} \right.
\label{eq:deff}
\end{equation}
So in order to compute the bias matrix for non-rotating beams,
only the $\Gamma$ symbols in Eqs.~(\ref{eq:BiasClebschEDS}) or
(\ref{eq:BiasWigdEDS}) have to be replaced by the above $f$
symbols. We thus arrive at the important conclusion that the
computation cost for estimating the bias for equal declination
scans is the {\em same} as that for non-rotating beams (only the
precomputed $\Gamma$ coefficients for that scan strategy would be
required). It is also reasonable to expect that the computation
cost would be of the same order even for a broader class of scan
strategies (e.g., as in case~I) which can provide reasonable
approximations to the real scan strategies.

\section{Limiting cases: Circular beam with full sky coverage \& incomplete sky coverage}
\label{sec:limit}

The analytic expressions for bias derived in the previous section
reduce to the known analytical results for non-uniform sky
coverage with circular beams studied in Hivon et al. (2002) and
our earlier results for the leading order effect of non-rotating
noncircular beams with full sky coverage (Mitra et al. 2004).

The special cases of circular beam and/or complete sky coverage limits
are readily recovered from our general expressions.

First, the simplest case of complete sky coverage with circular beam limit
can be obtained by substituting
$U_{lm} = \sqrt{4\pi} \delta_{l0}$ and $\B_{lm} = \delta_{m0}$.
We show in the Appendix~\ref{cbfs} that we get back the well known result
\begin{equation}
\nonumber A_{l l^\prime} \ = \ B_l^2 \, \delta_{ll'}.
\end{equation}

Hivon et al. (2002) formulated MASTER (Monte Carlo Apodized
Spherical Transform Estimator) method for the estimation of CMB
angular power spectrum from `cut' (incomplete) sky coverage for
circular beams. Substituting the circular beam limit [$\B_{lm} \delta_{m0}$] in the expression for bias matrix we recover the
MASTER circular beam result in Appendix~\ref{cbcs}:
\begin{equation}
A_{ll'} \ = \ B_l^2 \, \frac{2l' +1}{4\pi} \sum_{l''=|l-l'|}^{l+l'} \, (2l''+1)
\, \wjjj{l}{l'}{l''}{0}{0}{0}^2 \, {\cal{U}}_{l''},
\label{eq:master}
\end{equation}
where ${\cal{U}}_{\lpp} \equiv \sum_{\mpp=-\lpp}^{\lpp} |U_{\lpp\mpp}|^2/(2\lpp+1)$.

Finally, in Appendix~\ref{ncbfs}, we recover the general formula
for leading order correction with full sky coverage for
noncircular beams presented in Mitra et al. (2004). We substitute
$U_{lm} = \sqrt{4\pi} \delta_{l0}$ in the expression for the bias
matrix and get back
\begin{equation}
A_{l l'} \ = \ B_l^2 \, \frac{(2l'+1)}{4}
\sum_{m=-\rm{min}(l,l')}^{\rm{min}(l,l')} \left|
\sum_{m'=-l'}^{l'} \B_{l' m'} \, \int_{-1}^{1} d\cos\theta \,
d^l_{m0}(\theta) \, d^{l'}_{mm'}(\theta) \right|^2.
\end{equation}
Note that due to a somewhat different definition of bias matrix
in Mitra et al. (2004), for $\mathcal{C}_l \equiv
[l(l+1)/(8\pi^2)] C_l$, the results differ by a factor of
$[l'(l'+1)]/[l(l+1)]$ from  Eq.~(38) of Mitra et al. (2004).
Unfortunately, the  complex form of the final expression for
leading order correction to bias matrix with non-rotating beams
presented in Eq.~(43) of Mitra et al. (2004), does not allow an
explicit comparison of the results term by term.


\section{Numerical Implementation}\label{num}

The main motivation for deriving the analytic results is to evade
the computational cost and the time taken to estimate the bias
using end to end simulations. The present work takes us one step
ahead. The analysis framework we have developed can now estimate
the effect of non-circular beams including the effect of
non-uniform sky coverage for any given scan strategy. However,
the numerical evaluation of the algebraic expression for the bias
matrix derived in section~\ref{form} is also a computational
challenge. As discussed in section~\ref{form}, for a broad class
of scan strategies, which often provide good approximations to
real scan strategies (e.g., WMAP \cite{hin_wmap06}), the
computational cost for bias estimation can be significantly
reduced to the same order as needed for equal declination scan
strategies. More importantly, in situations where a
iso-declination scan strategy only provides a crude approximation
(e.g., Planck), the resulting bias estimate can be useful to
judge the severity of the asymmetric beam effect and, hence, to
decide at which level of rigor the problem needs to be addressed
in the analysis pipeline development.

In this section we describe
the detailed computation scheme for equal declination scan
strategies and show that the computational cost is within reach.
We also suggest certain criteria for the choice of the masks in
order to reduce the computational burden. Though the
implementation of our analysis can be computationally expensive
in the most general cases, even then the importance of this
method can not be underestimated. This analysis can illuminate
several ``shortcuts'' in the end-to-end simulations to reduce the
computational cost; furthermore, it has the potential to
eventually replace the end-to-end simulation entirely.

\subsection{Fast computation of the bias matrix}
\label{subsec:factComp}

Our scheme for fast computation of the bias is based on the
alternate form of the bias expressed by
Eq.~(\ref{eq:BiasWigdEDS}). Possibility of fast computation of
the form given by the combination of Eqs.~(\ref{eq:gen}) and
(\ref{eq:JClebsch}) is being considered.

The final analytic form of the bias matrix for equal declination
scans, given by Eq.~(\ref{eq:BiasWigdEDS}), contains infinite
summations. These summations have to be truncated for given
accuracy goals using reasonable physical insights. Let us denote
the ($l,m$) cut-offs for the mask and beam by ($l_{\rm{mask}},
m_{\rm{mask}}$) and ($l_{\rm{beam}}, m_{\rm{beam}}$) respectively.
The choice of the numerical values for these cut-offs will be
provided in the numerical results section.

Computation of the final expression by naively implementing the
analytic expression given by Eq.~(\ref{eq:BiasWigdEDS}) is
expensive. Three major innovations have been introduced in order
to numerically evaluate the bias matrix:

\begin{enumerate}
\item We used a smart implementation of the hierarchical summations
to reduce the computational cost by a few orders of magnitude.
To calculate three coupled loops of the form
\begin{equation}
S = \sum_{i=1}^{N} \sum_{j=1}^{N} \sum_{k=1}^{N} f(i+j+k),
\end{equation}
naively $N^3$ operations seem necessary. However, if we
calculate the summation in the following order:
\begin{eqnarray}
V(m) &:=& \sum_{k=1}^N f(m+k); \ \ \ m=1,2,\ldots,2N\\
S &=& \sum_{i=1}^{N} \sum_{j=1}^{N} V(i+j)
\end{eqnarray}
we effectively require just $2N^2 + N^2 = 3N^2$ operations. The
computational gain is $N/3$. For $N=3000$, this factor is 1000.
This example is for a very simple case where all the summations
have the same limits, but clearly this can be extended to the
case of summations with unequal limits and match our analysis
(See Appendix~\ref{app:fastCMBBias} for details). Here the
summations within the modulus symbols in (\ref{eq:BiasWigdEDS})
(that is for each set of $l,l',m,n$) are computed in three stages:
\begin{itemize}
\item Step I:
\begin{equation}
V^1(N) \ = \ \sum_{M'=-l'}^{l'}
d^{l'}_{mM'}\left(\frac{\pi}{2}\right)
\sum_{m'=-m_{\rm{beam}}}^{m_{\rm{beam}}} \beta_{l'm'} \,
d^{l'}_{M'm'}\left(\frac{\pi}{2}\right) \,
\Gamma_{m'(M'+N)}[\rho(\theta)]
\end{equation}
$N$ runs from $-(l + m_{\rm{mask}})$ to $+(l + m_{\rm{mask}})$

\item Step II:
\begin{equation}
V^2(M'') \ = \ \sum_{M=-l}^{l} d^{l}_{nM}\left(\frac{\pi}{2}\right) \,
d^{l}_{M0}\left(\frac{\pi}{2}\right) \, V^1(M+M'')
\end{equation}

\item Step III:
\begin{eqnarray}
V^3 &=& \sum_{l'' =0}^{l_{\rm{mask}}} \sqrt{2l''+1} \, U_{l''(m-n)} \ \times \nonumber\\
&&\sum_{M''=-m_{\rm{mask}}}^{m_{\rm{mask}}}
d^{l''}_{(m-n)M''}\left(\frac{\pi}{2}\right)
d^{l''}_{M''0}\left(\frac{\pi}{2}\right) V^2(M'')
\label{eq:fastBiasStep}
\end{eqnarray}
\end{itemize}
For $l_{\rm{beam}} = l_{\rm{max}}$ the computation time with naive
implementation would scale as  $\sim (8/3) (2m_{\rm{mask}} +1) (2
m_{\rm{beam}} + 1) l_{\rm{max}}^5 l_{\rm{mask}}^2$, whereas the
above algorithm reduces the computation cost to $\sim
(4/3)(2m_{\rm{mask}}+1)(2m_{\rm{beam}}+1)l^5_{\rm{max}}$,
providing a speed-up factor of $\sim 2 l_{\rm{mask}}^2$.

Mildly non-circular beams, where the BDP $\beta_{lm}$ at each $l$
falls off rapidly with $m$, allows us to neglect $\beta_{lm}$ for
$m > m_{\rm{beam}}$. For most realistic beams, $m_{\rm{beam}} \sim
4$ is a sufficiently good approximation (Souradeep \& Ratra 2001)
and this cuts off the summation over BDP in the bias matrix
$A_{l\lp}$.

Soft, azimuthally apodized, masks where the coefficients $U_{lm}$
are small beyond $m>m_{\mathrm{mask}}$, similarly allows us to
truncate the sum involving $U_{lm}$. Moreover, it is useful to
smooth the mask in $l$, such the $U_{lm}$ die off rapidly for
$l>l_{\mathrm{mask}}$ too.

Clearly, small values of ${m_{\rm{beam}}}$ and ${m_{\rm{mask}}}$
lead to computational speed up. Detailed discussion on mask
making is shown in section \ref{sec:mask}.

\item The Wigner-$d$ functions with argument $\pi/2$ occur too frequently in
the above evaluation. So one possibility to reduce computation cost was to
pre-compute all the Wigner-$d$ coefficients $d^l_{mn}(\pi/2)$ at once. But
for $l \sim 1000$ this scheme is limited by disk/memory storage and/or
program Input/Output (I/O) overhead.

However, we may observe that, in each step of computation
described in Eq.~(\ref{eq:fastBiasStep}) only one value of $l$
occurs in the $d$ symbols with the same argument $\pi/2$. Hence
we use an efficient recursive routine presented in Risbo (1996)
that generates all the $d^l_{mn}(\pi/2)$ at once for a given
value of $l$. This allows us to compute the Wigner-$d$ symbols
efficiently and use them as constant coefficients at each step
without any significant I/O limited operations.

\item There are several symmetries involved in the spherical
harmonic transforms and the Wigner-$d$ symbols, which could be
utilized to get more than an order of magnitude reduction in the
computation time.

\item Finally, we know that the bias matrix is not far off from diagonal, because
the beams are mildly non-circular (see the plots of  bias matrices in \cite{mitra04}).
So we need not compute all the elements of the
bias matrix. Rather, a diagonal band (could be of triangular shape) of
average ``thickness'' $\Delta l$ can be used to calculate the $C_l$ estimation error
with a fairly high accuracy. This will give an additional speed-up factor of $\sim l/\Delta l$.

\end{enumerate}

While speeding up the numerical implementation of the above
analysis is under progress, the estimate of computational
requirement for the basic code for $l_{\rm{max}} = 3000$ is quite
promising, well within the computing resources available to the
CMB community (see sec.~\ref{sec:biastime} for details).

It is also interesting to note here that for different models of the same beam,
which would at most differ at a highly constrained band in the harmonic space,
the bias computation needs to be repeated only for those harmonic components.
This would save large amount of computation. This advantage may not be available
to pixel based end to end Monte Carlo simulation methods for estimating bias.


\subsection{Constructing azimuthally apodized masks }\label{sec:mask}

The temperature anisotropies observed by any detector are
combinations of CMB as well as foreground. The dominant
contribution of foreground arises from the galactic plane. While
methods of foreground removal using multi-wavelength observations
exist, there is significant residual along the galactic plane to
require masking of that region prior to cosmological power
spectrum estimation.  The mask is designed to remove the effect
of regions of excessive galactic emission and spots around strong
extragalactic radio sources (Bennett et al. 2003b, Saha et al.
2006).

\begin{figure*}
\begin{center}
\vspace{0.5cm}
\includegraphics[width=0.3\textwidth,angle=90]{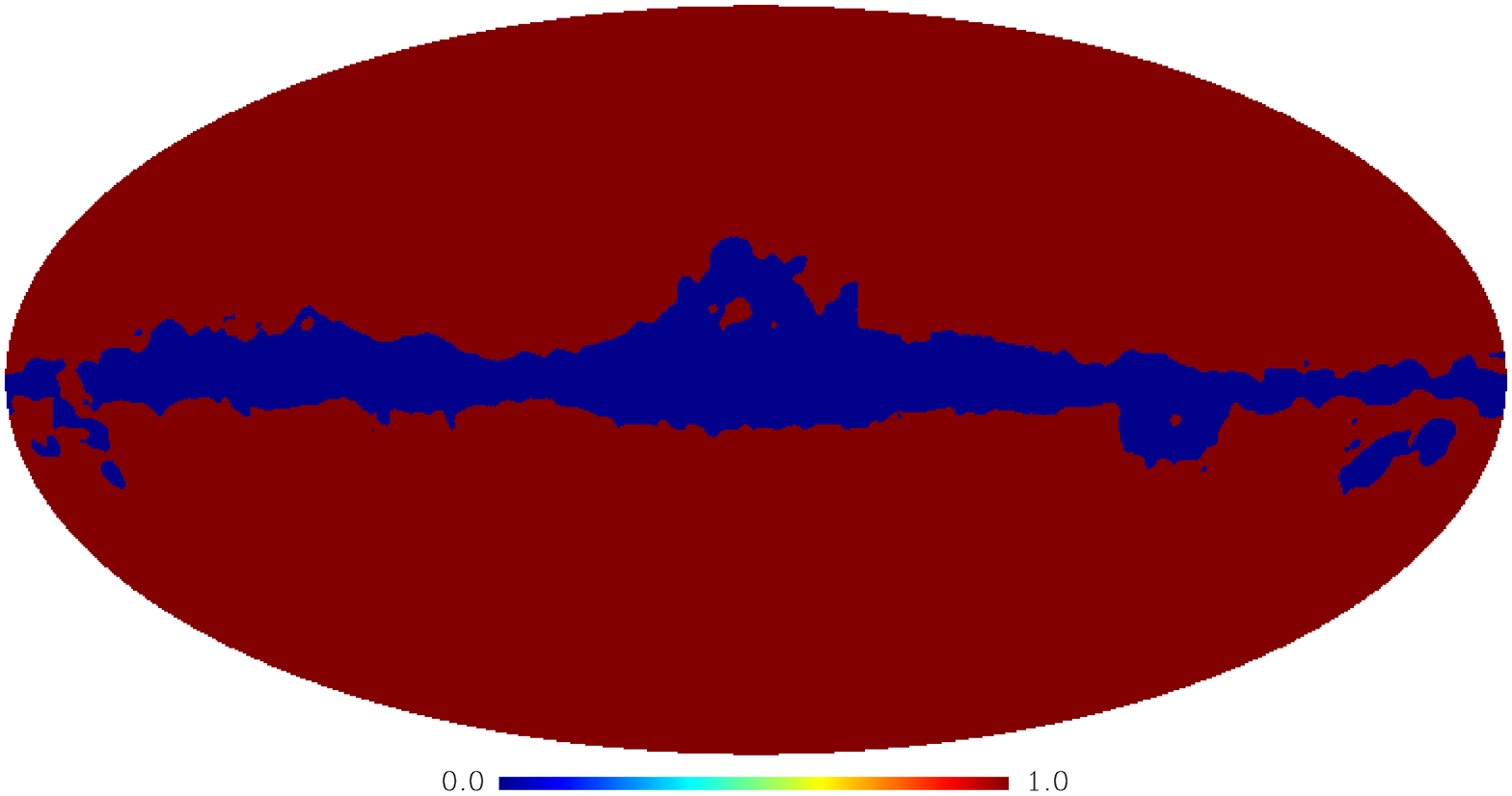}
\includegraphics[width=0.3\textwidth,angle=90]{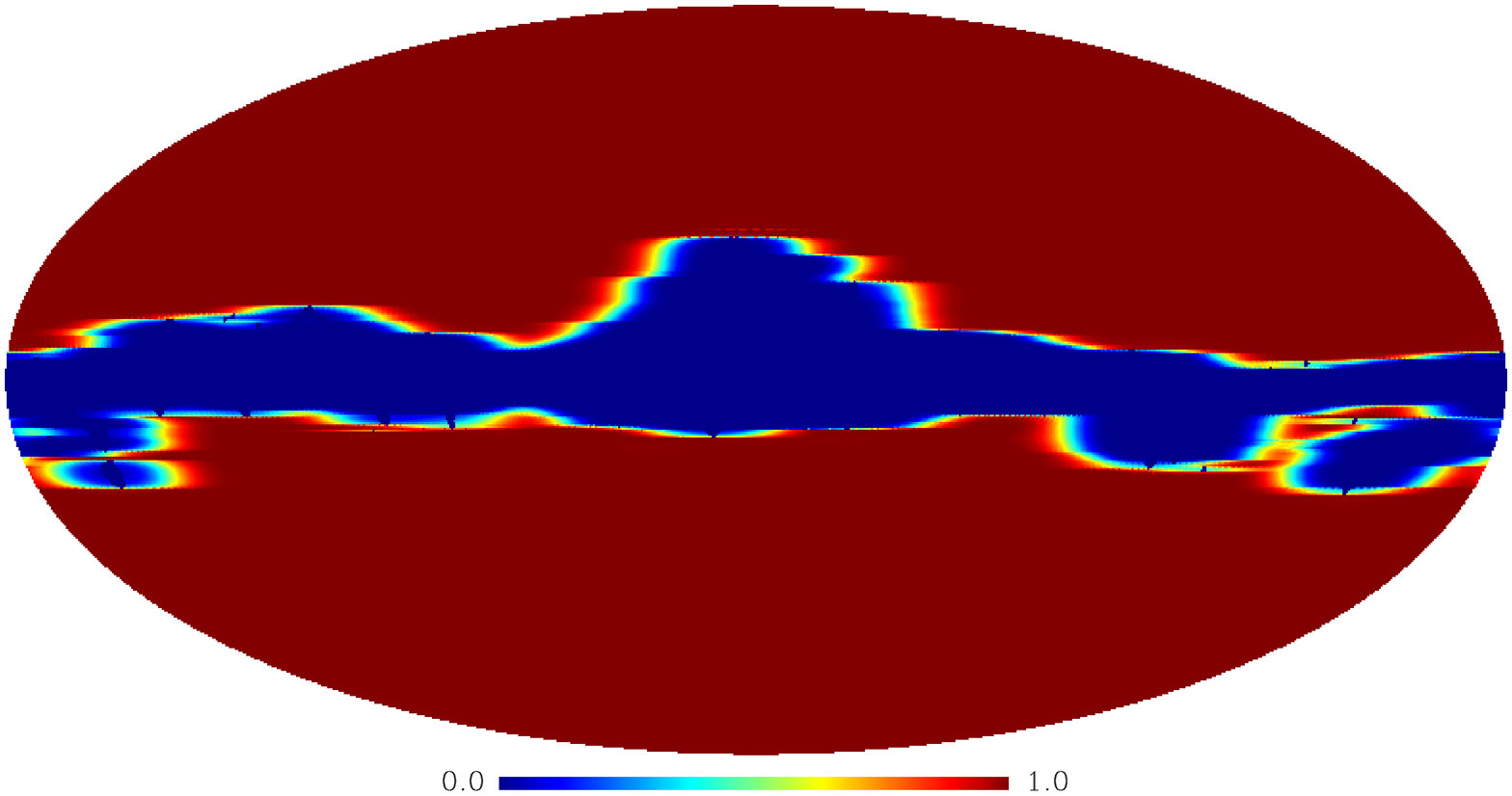}
\caption{The KP2 original mask (left) and the azimuthally
smoothed mask reconstructed from the KP2 mask (right).}
\label{fig:mask}
\end{center}
\end{figure*}

The effect of masked map on the angular power spectrum estimation
has been described in the literature (Hivon et al. 2002).
However, as we have shown, the effect of cut sky becomes
nontrivial for a non-circular beam function. The sum over all the
$U_{lm}$ modes of the mask is responsible for the large
additional computation cost. In particular, our computational
cost estimate in the previous subsection shows that a mask that
allows us to choose a modest value of $m_{\rm{mask}}$ leads to a
proportionally smaller computational cost. A mask whose transform
is such that power in high $m$ modes for a given $l$ is
suppressed would clearly serve this purpose (Souradeep et al.,
2006).

\begin{figure*}
\includegraphics[width=.33\textwidth,angle =-90]{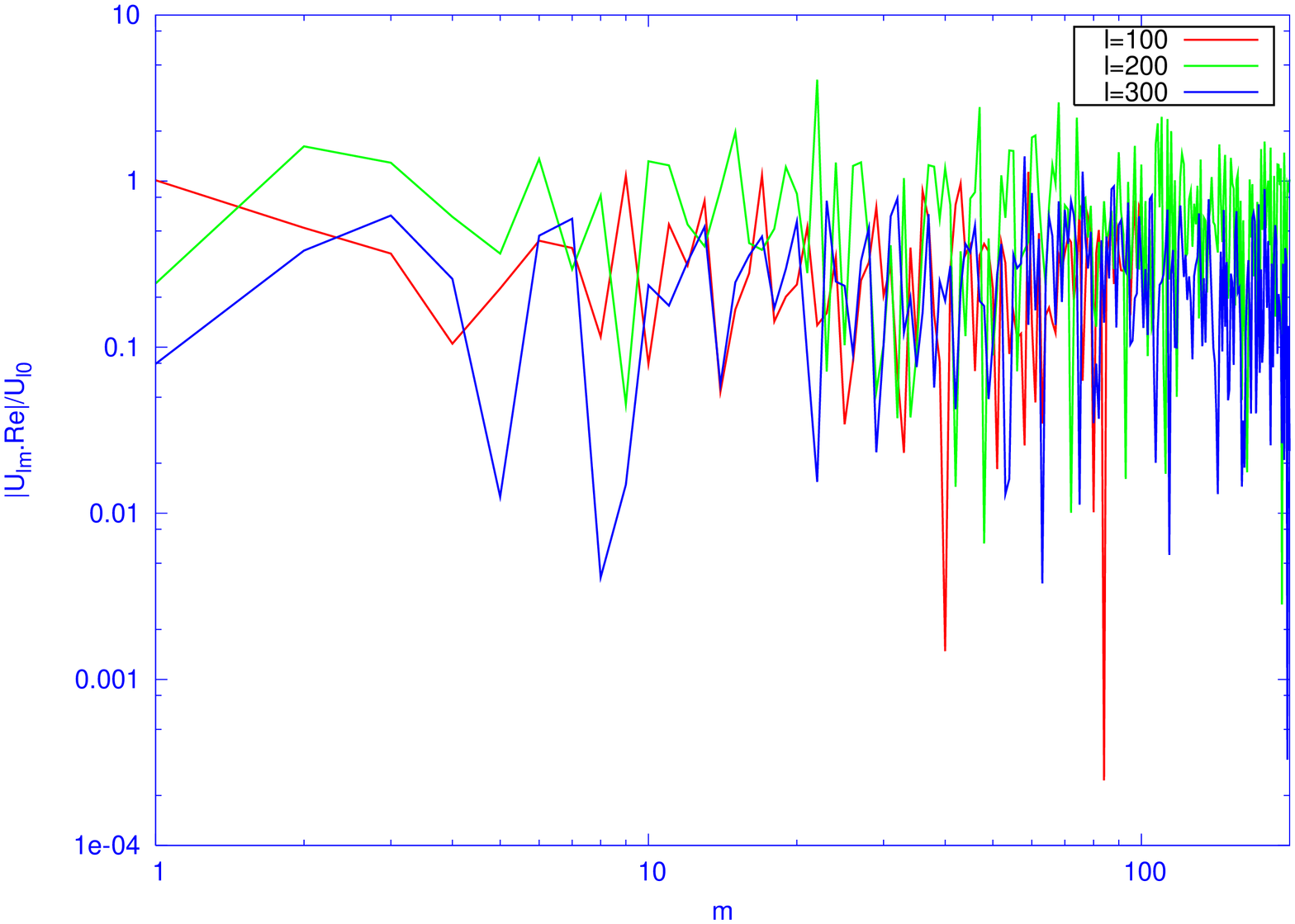}
\includegraphics[width=.33\textwidth,angle =-90]{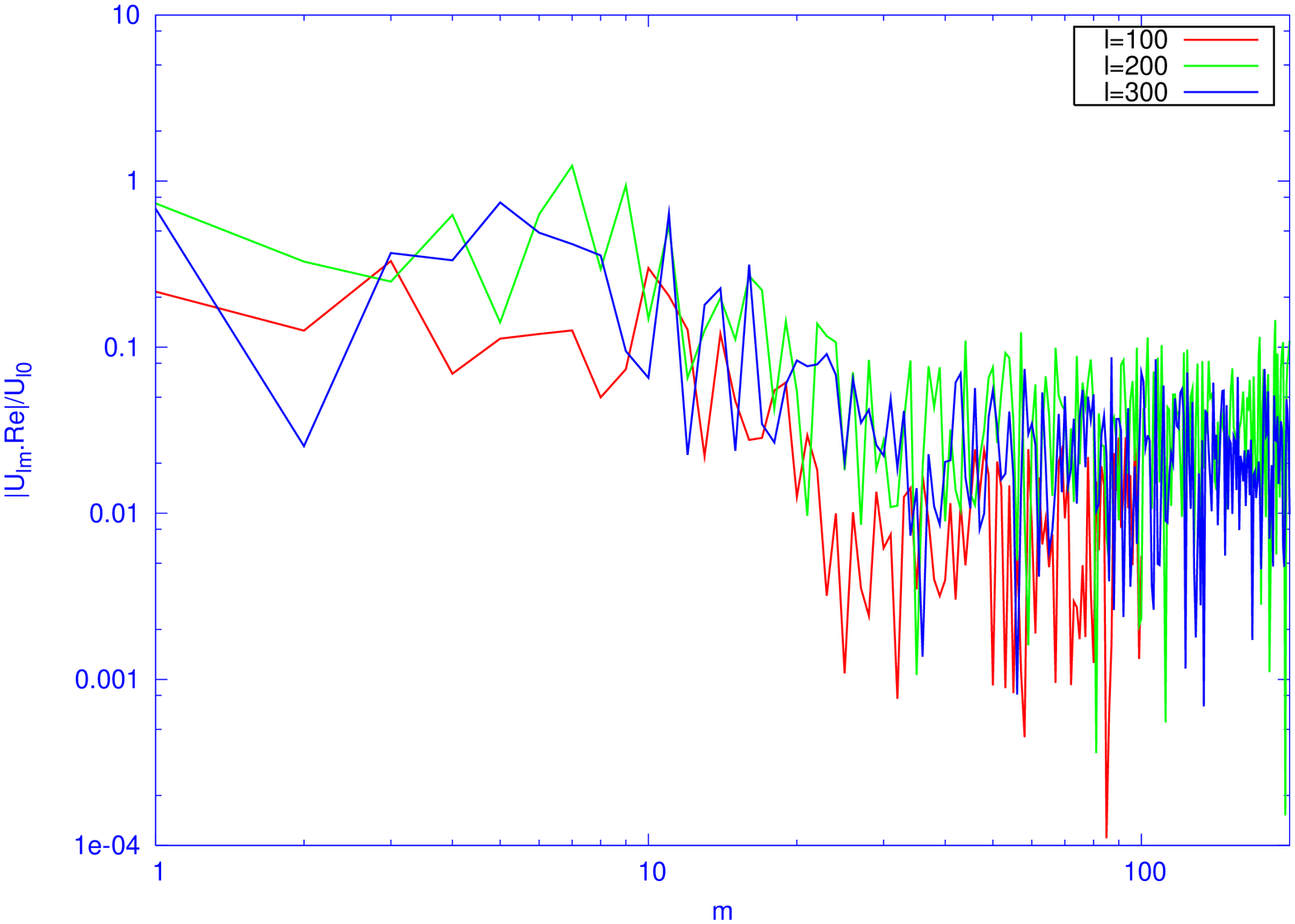}
\caption{$U_{lm}$s of the original mask KP2 (left) and the mask Kp2$^\prime$ (right).}
\label{fig:ulm}
\end{figure*}

To achieve this we propose a possible method for generating an
azimuthally-apodized version of a given mask~\footnote{For
concreteness, we consider the example of Kp2 mask used in the WMAP
analysis Here, for simplicity we do not consider the excised point
sources. As this mask has also 0.6 degree radius cut around 208
locations of the point sources we first fill them up, except few, very
near the galactic plane.} as outlined in the following steps:

\begin{enumerate}

\item
We compute spherical harmonic coefficients $U_{lm} $ of the
original mask $U^o(\hat q)$. We directly suppress the power at
high $m$, by re-scaling
\begin{equation}
    U^\prime _{lm}=\exp{(-m*m/[\alpha \, {m_{\rm{mask}}}^2])}*U_{lm}
\end{equation}
corresponding to smoothing the mask along the azimuthal direction.
Where $\alpha$ determines the extent to which power is suppressed
at the given cut-off value of $m_{\rm{mask}}$.

\item The re-scaled $U_{lm}$ are transformed to make an auxiliary mask
$U^\prime(\hat q)$.

\item However, it is clear that mask $U^\prime(\hat q)$ would allow power
from contaminated regions. We should ensure that all regions where
$U^o(\hat q)=0$ remain zero. A simple way to do that would be to multiply
$U^\prime(\hat q)$ with $U^o(\hat q)$ in the pixel space. Hence, we
define the final apodized mask as

\begin{equation}
U^a(\hat q) = [U^\prime(\hat q)]^s \, U^o(\hat q),
\end{equation}
where $s$ is a sufficiently large number ($\alpha=1$, $m_{\rm
{mask}}=10$ and $s= 12$ in our example shown in figures
\ref{fig:mask} \& \ref{fig:ulm}) to ensure that the edges of the
final mask are smoothed to the required level. To be more
explicit, multiplying $U'$ with $U_0$  would introduce steps in
the mask. Raising $U'$ to be a sufficiently high power ensures
that the amplitude of the step is reduced.

\end{enumerate}

The final mask obtained in this method is shown in the right
panel of figure \ref{fig:mask}.  We extract $U_{lm}$ from this
final mask $U^a(\hat q)$. We show the $U_{lm}$ of the original
mask KP2 and the azimuthally apodized mask Kp2$^a$ in figure
\ref{fig:ulm}. Clearly, the later show very rapid decrease of a
mode with $m$ for a given $l$. In the example shown here, we have
significant contribution only from the first 10-20 $m$ modes, for
a given $l$.  The $|U_{lm}|^2$ also dies down with $l$ allowing
us also to put a cut-off at $l =l_{\rm{mask}} \sim 100$.  A
reconstructed mask following this method has the advantage that
it reduces the computation time for bias matrix by a large factor.

\subsection{Estimate of time for calculating the Bias matrix}
\label{sec:biastime}

We have (semi)empirically estimated the CPU time required by our
codes for equal declination scan strategies with apodized masks.
Since the computation cost for equal declination scan strategies
is the \emph{same} as that for non-rotating beams, we use the
latter for simplicity.

\begin{figure}
\centering
\includegraphics[height=0.7\textwidth, angle=-90]{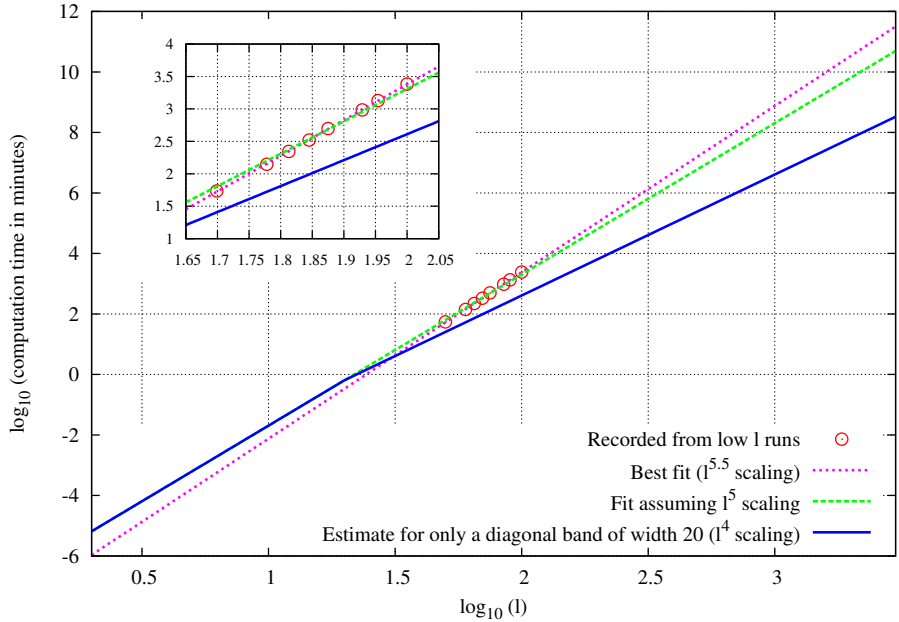}
\caption{Circled points are the datapoints of the sample time for running the code
at different multipoles. The dotted line is the best fit curve. The dashed line is the
theoretically expected estimate for the full bias matrix. The solid line below is the
estimated computation will be required if we consider only a diagonal
band of width $\Delta l$ = 20 elements in the bias matrix elements where
the effect of non-circular beams is significant.}
\label{timing}
\end{figure}

We ran our codes for maximum multipoles $l_{\rm{max}}$ of 50, 60,
65, 70, 75, 85, 90 \& 100 in a 8 processor (4$\times$ AMD Opteron
Dual core at 2.6 Ghz) node and having 16 GB of RAM. The recorded
computation times are shown in the log - log plot in
figure~\ref{timing} (circled points). We fitted these points with
linear functions of the form $y(x)=\alpha x +\beta $. The best
fit is obtained for $\beta=-7.6$ and $\alpha=5.5$ (dotted line),
which indicates that the observed computation time scales as
$l^{5.5}$, whereas algorithmically we show that it should be
$l^5$. While the theoretical prediction (dashed line) is also
quite close to the observed values, the discrepancy is perhaps
related to the simplistic implementation we have used in our
codes and this difference should reduce in near future as the
code evolves.

Further reduction of computation is conceivable in the future:
\renewcommand{\labelenumi}{\Roman{enumi}.}
\begin{enumerate}
\item If we take into account that the bias matrix is {\it most
significant} only along a narrow strip (approximately $\Delta l$
= 20 elements wide on average and not necessarily uniform - could
be wedge shaped) around the diagonal of the bias matrix, the time
estimate reduces by another power of $l$.
The solid line in Figure~\ref{timing} refers to the plot of
estimated computation time against multipole in the log scale,
with the slope reduced by 1 (i.e. 5-1=4). When $l_{\rm{max}} \le
20$, of course, there is no reduction in computation cost, which
is shown by the first part of the solid curve.

\item Using the symmetries of the Wigner-d symbols, it is possible to
reducing the computation cost by more than an order of magnitude.

\end{enumerate}

If all the above modifications are implemented, the bias calculation is
expected to be feasible in quite reasonable time with the computing
facilities available/dedicated to the CMB community. For example, with
1000 dual core CPUs the bias matrix for $l_{\rm{max}} = 3000$,
$m_{\rm{mask}} = 20$, $m_{\rm{beam}} = 2$ and $\Delta l = 20$ should be
computed in 8 weeks time.
Note that, this is a conservative estimate. With advanced
implementation of the numerics having exact algorithmic scaling,
the computation time can reduce by as much as one order of
magnitude.


\section{Discussions}
\label{disc}

The inclusion of the effect of noncircular beam leads to major
complications at every stage of the data analysis pipeline.  The
extent to which the non-circularity affects the step of going
from the time-stream data to sky map is also very sensitive to
the scan-strategy. The beam now has an orientation with respect
to the scan path that can potentially vary along the path. This
implies that the beam function is inherently time dependent and
difficult to deconvolve.

In our present work, we have extended our analytic approach for
estimating the leading order bias due to noncircular experimental
beam on the angular power spectrum ${C}_l$ of CMB anisotropy -
the analytical framework now includes the effect of incomplete
sky coverage and it is no more limited to only the leading order
correction. It can also incorporate the case of equal declination
scans without demanding any change in the codes or the
computation cost. Though the numerical implementation does not
include the effect of most general scan strategy, the formalism
presented in this work is valid for estimating the full bias
correction using a (semi)analytic perturbative method that can
replace the computationally costly end-to-end simulation used for
current CMB experiments. This work also provides an analytic
framework to perform the simulation steps more efficiently.

Noncircular beam effects can be modeled into the covariance
functions in approaches related to maximum likelihood estimation
(Tegmark 1997, Bond et al. 1998) and can also be included in the
Harmonic ring (Challinor et al. 2002) and ring-torus estimators
(Wandelt \& Hansen 2003). However, all these methods are
computationally prohibitive for high resolution maps and, at
present, the computationally economical approach of using a
Pseudo-$C_l$ estimator appears to be a viable option for
extracting the power spectrum at high multipoles (Efstathiou
2004). The Pseudo-$C_l$ estimates have to be corrected for the
systematic biases.  While considerable attention has been devoted
to the effects of incomplete/non-uniform sky coverage, no
comprehensive or systematic approach is available for noncircular
beam.  The high sensitivity, `full' (large) sky observation from
space (long duration balloon) missions have alleviated the effect
of incomplete sky coverage and other systematic effects such as
the one we consider here have gained more significance.
Non-uniform coverage, in particular, the galactic masks affect
only CMB power estimation at the low multipoles. The analysis
accompanying the recent second data release from WMAP uses the
hybrid strategy (Efstathiou 2004) where the power spectrum at low
multipoles is estimated using optimal Maximum Likelihood methods
and Pseudo-$C_l$ are used for large multipoles Hinshaw et al.
2007, Spergel et al. 2007).

The noncircular beam is an effect that dominates at large $l$
beyond the the inverse beam width (Mitra et al. 2004). For high
resolution experiments, the optimal maximum likelihood methods
which can account for non-circular beam functions are
computationally prohibitive.  In implementing the Pseudo-$C_l$
estimation, we have included both the non-rotating noncircular
beam effect and the effect of non-uniform sky coverage. Our
preliminary estimate shows that the computation cost for
$l_{\rm{max}} \sim 3000$ is within reach. Furthermore, equal
declination scan strategies can be trivially included in this
implementation. Our work provides a convenient approach for
estimating the magnitude of these effects in terms of the leading
order deviations from a circular beam and azimuthally symmetric
mask. The perturbative approach is very efficient. For most CMB
experiments the leading few orders capture most of the effect of
beam non-circularity (Souradeep \& Ratra 2001). Our results
highlight the advantage of azimuthally smoothed masks (mild
deviations from azimuthal symmetry) in reducing computational
costs. This process is more efficient as compared to the
isotropic apodization of masks (Efstathiou 2006), that suffers a
lot more information loss at the edges. The numerical
implementation of our method can readily accommodate the case
when pixels are revisited by the beam with different
orientations. Evaluating the realistic bias and error-covariance
for a specific CMB experiment with noncircular beams would
require numerical evaluation of the general expressions for
$A_{l\lp}$ using real scan strategy and account for inhomogeneous
noise and sky coverage, the latter part of which has been
addressed in this present work.

It is worthwhile to note in passing that that the angular power
$C_l$ contains all the information of Gaussian CMB anisotropy
only under the assumption of statistical isotropy.  Gaussian CMB
anisotropy map measured with a non-circular beam corresponds to
an underlying correlation function that violates statistical
isotropy. In this case, the extra information present may be
measurable using, for example, the bipolar power spectrum (Hajian
\& Souradeep 2003, Hajian et al. 2004, Hajian \& Souradeep 2006,
Basak et al. 2006). Even when the beam is circular the scanning
pattern itself is expected to cause a breakdown of statistical
isotropy of the measured CMB anisotropy (Hivon et al. 2002). For
a non-circular beam, this effect could be much more pronounced
and, perhaps, presents an interesting avenue of future study.
Accounting for the noncircular beam may also be crucial in the
context of non-gaussianity measurements. Saha et al. (2008)
developed a procedure to estimate the CMB power spectrum directly
from the raw CMB data without the need of foreground templates. A
CMB map made with non-circular beam but analyzed assuming a
circular beam could induce higher order correlations that could be
misinterpreted as a primordial non-gaussian signal. This is
perhaps more critical for methods that first deconvolve the
primordial perturbation from the CMB maps (Yadav \& Wandelt, 2005
and Yadav \& Wandelt 2008) since differences between the actual
and circular beam approximations could get amplified and
propagated through the step of deconvolution.

In addition to temperature fluctuations, the CMB photons coming
from different directions have a random, linear polarization. The
polarization of CMB can be decomposed into $E$ part with even
parity and $B$ part with odd parity.  Besides the angular spectrum
$C_l^{TT}$, the CMB polarization provides three additional
spectra, $C_l^{TE}$, $C_l^{EE}$ and $C_l^{BB}$ which are
invariant under parity transformations. The level of polarization
of the CMB being about a tenth of the temperature fluctuation, it
is only very recently that the angular power spectrum of CMB
polarization field has been detected. The Degree Angular Scale
Interferometer (DASI) has measured the CMB polarization spectrum
over limited band of angular scales in late 2002 (Kovac et al.
2002). The DASI experiment recently published 3-year results of
much refined measurements (Lietch et al. 2005). More recently,
the BOOMERanG collaboration reported new measurements of CMB
anisotropy and polarization spectra (Piacentini et al. 2006,
MacTavish et al. 2006). The WMAP mission has also measured CMB
polarization spectra (Kogut et al. 2003, Page et al. 2007).
Correcting for the systematic effects of a non-circular beam for
the polarization spectra is expected to become important.
Extending this work to the case CMB polarization is another line
of activity we plan to undertake in the near future.

In summary, we have presented a perturbation framework to compute
the effect of non-circular beam function on the estimation of
power spectrum of CMB anisotropy taking into account the effect
of a non-uniform sky coverage (e.g., galactic mask). We not only
present the most general expression including non-uniform sky
coverage as well as a noncircular beam that can be numerically
evaluated but also provide elegant analytic results in
interesting limits which are useful for gathering insights to
efficiently analyze data. As CMB experiments strive to measure
the angular power spectrum with increasing accuracy and
resolution, the work provides a stepping stone to address a rather
complicated systematic effect of noncircular beam functions.

\section*{Acknowledgments}

SM would like to thank Council of Scientific and Industrial
Research (India) for supporting his research.
Part of the writing of this paper was
carried out at the Jet Propulsion Laboratory, California Institute of Technology,
under a contract with the National Aeronautics and Space Administration.
We thank Kris Gorski, Jeff Jewel \& Ben Wandelt for providing the reference and
a code for computing Wigner-$d$ functions. We thank Olivier Dore
and Mike Nolta for providing us with the data files of the
non-circular beam correction estimated by the WMAP team. We
acknowledge the fruitful discussions with Francois Bouchet, Simon
Prunet and Charles Lawrence. Computations were carried out at the
HPC facility available at IUCAA.


\appendix

\section{Evaluation of $J^{l\lpp\lp}_{n\mpp mm^\prime}$ }

We first evaluate $J^{l\lpp\lp}_{n\mpp mm^\prime}$ in
Eq.~(\ref{eq:gen}) using use the Clebsch-Gordon coefficients and
sinusoidal expansion of the Wigner-$d$ functions for the general
case and for the case of equal declination scans. But for
efficient numerical implementation in the case of non-rotating
beams, we employ a different strategy using sinusoidal expansion
of the Wigner-$d$ functions that involve only the
$d^{l}_{mm'}(\pi/2)$ symbols.

\subsection{Approach I: Using Clebsch-Gordon series and expansion of
Wigner-$d$}
\label{appendix:clebs}

Putting Eq~(\ref{CGSY}), (\ref{YstarY}), (\ref{YD}),
(\ref{CGSD}) \& (\ref{eq:wigdexp}) in Eq~(\ref{eq:defJ}) we get
\begin{eqnarray}
J^{l\lpp\lp}_{n\mpp mm^\prime} &:=&
\int_{4\pi} d\Omega_{\mathbf{\hat{q}}} \,
Y_{ln}(\mathbf{\hat{q}}) \, Y_{\lpp\mpp}(\mathbf{\hat{q}}) \,
D_{mm^\prime}^{l^\prime}(\mathbf{\hat{q}},\rho(\mathbf{\hat{q}}))\\
&=& \sum_{L=|l-\lpp|}^{l+\lpp}
\sqrt{\frac{(2l+1)(2\lpp+1)}{4\pi(2L+1)}} \, C^{L0}_{l0\lpp 0} \,
C^{L(n+\mpp)}_{ln\lpp \mpp} \ \times \nonumber\\
&& \ \ \ \int_{4\pi}
d\Omega_{\mathbf{\hat{q}}} \, Y_{L(n+\mpp)}(\mathbf{\hat{q}})
D_{mm^\prime}^{l^\prime}(\mathbf{\hat{q}},\rho(\mathbf{\hat{q}}))
\nonumber\\
&=& (-1)^{n+\mpp} \frac{\sqrt{(2l+1)(2\lpp+1)}}{4\pi}
\times \nonumber\\ && \sum_{L=|l-\lpp|}^{l+\lpp}  C^{L0}_{l0\lpp
0} C^{L(n+\mpp)}_{ln\lpp \mpp} \int_{4\pi}
d\Omega_{\mathbf{\hat{q}}}
D^{L}_{(-n-\mpp)0}(\mathbf{\hat{q}},\rho(\mathbf{\hat{q}}))
D_{mm^\prime}^{l^\prime}(\mathbf{\hat{q}},\rho(\mathbf{\hat{q}}))
\nonumber\\ &=& (-1)^{n+\mpp}
\frac{\sqrt{(2l+1)(2\lpp+1)}}{4\pi}\sum_{L=|l-\lpp|}^{l+\lpp}
C^{L0}_{l0\lpp 0} \, C^{L(n+\mpp)}_{ln\lpp \mpp}
\times \nonumber\\ &&  \sum_{\Lp=|L-\lp|}^{L+\lp}
C^{\Lp(m-n-\mpp)}_{L(-n-\mpp)\lp m} \, C^{\Lp m^\p}_{L0\lp m^\p} \,
\chi^{\Lp}_{(m-n-\mpp)m^\p}[\rho(\q)],
\nonumber
\end{eqnarray}
where
\begin{equation}
\chi^{l}_{mm'}[\rho(\q)] \ := \ \int_{4\pi} d\Omega_{\mathbf{\hat{q}}} \,
D^{l}_{mm^\p}(\mathbf{\hat{q}},\rho(\mathbf{\hat{q}})).
\end{equation}

The above algebra gives us the most general expression for bias.
But, as discussed in the text, special cases, which are
computationally beneficial, often provide good approximation to
the real scan strategies. For equal declination scan strategies,
$\rho(\q) \equiv \rho(\theta)$,
\begin{equation}
\chi^{l}_{mm'}[\rho(\theta)] \ = \ \int_0^{2\pi} d\phi \,
e^{-im\phi} \int_0^\pi d\theta \, d^l_{mm'}(\theta) \, e^{-i m'
\rho(\theta)} \ = \ 2\pi \, \delta_{m0} \, \int_0^\pi d\theta \,
d^l_{0m'}(\theta) \, e^{-i m' \rho(\theta)}.
\end{equation}
Hence, in that case,
\begin{eqnarray}
J^{ll''l'}_{nm'' mm'} &=&  \delta_{m''(m-n)} \, (-1)^{m}
\frac{\sqrt{(2l+1)(2\lpp+1)}}{4\pi}\sum_{L=|l-\lpp|}^{l+\lpp}
C^{L0}_{l0\lpp 0} \, C^{Lm}_{ln\lpp \mpp}
\times \nonumber\\ &&  \sum_{\Lp=|L-\lp|}^{L+\lp}
C^{\Lp 0}_{L -m \lp m} \, C^{\Lp m^\p}_{L0\lp m^\p} \,
\chi^{\Lp}_{0m'}[\rho(\theta)]
\end{eqnarray}
and the final expression for the bias matrix becomes
\begin{eqnarray}
A_{ll'} &=& B_l^2 \, \frac{(2\lp+1)}{64\pi^3} \sum_{n=-l}^{l}
\sum_{m=-l^\prime}^{l^\prime} \left| \sum_{\lpp =0}^{\infty}
\sqrt{2\lpp+1} \, U_{\lpp(m-n)} \right. \ \times\nonumber\\
&&\sum_{L=|l-\lpp|}^{l+\lpp} \left.
C^{L0}_{l0\lpp 0} \, C^{Lm}_{ln\lpp(m-n)}  \sum_{\Lp=|L-\lp|}^{L+\lp}
C^{\Lp 0}_{L-m\lp m}
\sum_{m'=-l'}^{l'} \beta_{l' m'} \, C^{L' m'}_{L0l'm'} \, \chi^{\Lp}_{0m'}[\rho(\theta)] \right|^2.
\label{eq:biasMatClebEquiDecl}
\end{eqnarray}

The coefficients $\chi^{l}_{mm'}$ could also be expressed in an
alternative (insightful) form using Eq.~(\ref{eq:wigdexp}) as:
\begin{equation}
\chi^{l}_{mm'}[\rho(\q)] \ = \ i^{m+m'} \sum_{N=-l}^{l} (-1)^N \, d^{l}_{mN}
\left(\frac{\pi}{2}\right) \,
d^{l}_{Nm'}\left(\frac{\pi}{2}\right) \int_{4\pi}
d\Omega_{\mathbf{\hat{q}}} \, e^{-im\phi} \, e^{iN\theta} \,
e^{-im'\rho(\mathbf{\hat{q}})}.\nonumber
\end{equation}
Then for equal declination scan strategies one gets
\begin{equation}
\chi^{\Lp}_{0m'}[\rho(\theta)] \ = \ 2\pi \sum_{N=-l}^{l} d^{l}_{mN}\left(\frac{\pi}{2}\right) \,
d^{l}_{Nm'}\left(\frac{\pi}{2}\right) \, \Gamma_{m'N}[\rho(\theta)],
\end{equation}
where [Eq.~(\ref{eq:defGamma})]
\begin{equation}
\Gamma_{m'N}[\rho(\theta)] \ := \ i^{m^\p} \, (-1)^N \int_0^\pi
\sin\theta d\theta \, e^{iN\theta} \, e^{-i m' \rho(\theta)},
\end{equation}
and the $J$ coefficients become
\begin{eqnarray}
J^{l\lpp\lp}_{n\mpp mm^\prime} &=& (-1)^{n+\mpp} \,
\delta_{\mpp(m-n)} \, \frac{\sqrt{(2l+1)(2\lpp+1)}}{2} \ \times
\nonumber\\
&&\sum_{L=|l-\lpp|}^{l+\lpp} C^{L0}_{l0\lpp 0} \,
C^{L(n+\mpp)}_{ln\lpp \mpp} \sum_{\Lp=|L-\lp|}^{L+\lp}
C^{\Lp(m-n-\mpp)}_{L(-n-\mpp)\lp m} \, C^{\Lp m^\p}_{L0\lp m^\p} \
\times\nonumber\\
&&\sum_{N=-\Lp}^{\Lp} \, d^{\Lp}_{0N}\left(\frac{\pi}{2}\right) \,
d^{\Lp}_{Nm^\p}\left(\frac{\pi}{2}\right) \, \Gamma_{m'N}[\rho(\theta)]. \label{eq:JClebschEDS}
\end{eqnarray}

\subsection{Approach II: Using sinusoidal expansion of
Wigner-$d$}\label{appendix:wigd}
We start by plugging in Eq.~(\ref{eq:wigdexp}) in Eq.~(\ref{eq:defJ})
\begin{eqnarray}
\nonumber
J^{l\lpp\lp}_{n\mpp mm^\prime} &:=& \int_{4\pi}
d\Omega_{\mathbf{\hat{q}}} \, Y_{ln}(\mathbf{\hat{q}}) \,
Y_{\lpp\mpp}(\mathbf{\hat{q}}) \,
D_{mm^\prime}^{l^\prime}(\mathbf{\hat{q}},\rho(\mathbf{\hat{q}}))\\
&=& \frac{\sqrt{(2l+1)(2\lpp+1)}}{4\pi} \int_0^{2\pi} d\phi \,
e^{i(n+\mpp-m)\phi} \ \times\nonumber\\
&&\int_0^\pi \sin\theta d\theta \, d_{n0}^l(\theta) \,
d_{\mpp 0}^\lpp(\theta) \, d_{mm^\prime}^{l^\prime}(\theta) \, e^{-i
m^\prime \rho(\q)} \nonumber\\
&=& \frac{\sqrt{(2l+1)(2\lpp+1)}}{4\pi} \, i^{n+m+m''}
\sum_{M=-l}^{l} d^{l}_{nM}\left(\frac{\pi}{2}\right) \,
d^{l}_{M0}\left(\frac{\pi}{2}\right) \times
\nonumber\\
&&\sum_{\Mpp=-\lpp}^{\lpp}
d^{\lpp}_{\mpp\Mpp}\left(\frac{\pi}{2}\right) \,
d^{\lpp}_{\Mpp0}\left(\frac{\pi}{2}\right) \sum_{M^\p=-\lp}^{\lp}
d^{\lp}_{mM^\p}\left(\frac{\pi}{2}\right) \, d^{\lp}_{M^\p
m^\p}\left(\frac{\pi}{2}\right) \times \nonumber\\
&& i^{m'} \, (-1)^{M+M''+M'} \int_0^{2\pi} d\phi \, e^{i(n+m''-m)\phi}
\int_0^\pi \sin\theta d\theta \, e^{i(M+M^\p+\Mpp)\theta} \, e^{-i
m^\prime \rho(\q)}. \label{eq:JWigdExplicit}
\end{eqnarray}

This is the expression for the bias in its full generality. For a
specified general form of $\rho(\q)$, we need to precompute the
integral
\begin{equation}
\Xi_{m_1 m_2 m_3} \ := \ (-1)^{m_1} \, i^{m_3} \int_0^{2\pi} d\phi \, e^{i m_2 \phi}
\int_0^\pi \sin\theta d\theta \, e^{i m_1 \theta} \, e^{-im_3 \rho(\q)}.
\end{equation}
For equal declination scans ($\rho(\q)=\rho(\theta)$)
\begin{equation}
\Xi_{m_1 m_2 m_3} \ = \ 2\pi \, \delta_{m_20} \, \Gamma_{m_3m_1}[\rho(\theta)],
\end{equation}
so the $J$ coefficients take the form
\begin{eqnarray}
J^{l\lpp\lp}_{n\mpp mm^\p} &=&  2\pi \, \delta_{(m-n)\mpp} \, (-1)^m\,
\frac{\sqrt{(2l+1)(2\lpp+1)}}{4\pi} \ \times \nonumber\\
&&  \sum_{M=-l}^{l} d^{l}_{nM}\left(\frac{\pi}{2}\right)
d^{l}_{M0}\left(\frac{\pi}{2}\right) \sum_{\Mpp=-\lpp}^{\lpp}
d^{\lpp}_{(m-n)\Mpp}\left(\frac{\pi}{2}\right)
d^{\lpp}_{\Mpp0}\left(\frac{\pi}{2}\right) \ \times \nonumber\\
&& \sum_{M^\p=-\lp}^{\lp}
d^{\lp}_{mM^\p}\left(\frac{\pi}{2}\right) \, d^{\lp}_{M^\p
m^\p}\left(\frac{\pi}{2}\right) \, \Gamma_{m^\p(M+M^\p+\Mpp)}[\rho(\theta)].
\label{eq:JWigdEDS}
\end{eqnarray}

\section{Bias for non-rotating beams}
\label{app:noRot}

As discussed in the text, the study of the bias computation for {\em
non-rotating beams} provide a framework that is computationally
equivalent to broader class of scan strategies. Hence we treat the case
of non-rotating beams with greater importance. In this case the $\Gamma$
symbols and hence the final expression can be expressed in a closed
form. The algebra has been worked out below.

Substituting $\rho(\theta) = 0$ in Eq.~(\ref{eq:JWigdExplicit})
one gets,
\begin{eqnarray}
J^{l\lpp\lp}_{n\mpp mm^\p} &=& \delta_{m''(m-n)} \, \frac{\sqrt{(2l+1)(2\lpp+1)}}{2}
\sum_{M=-l}^{l} d^{l}_{nM}\left(\frac{\pi}{2}\right) \,
d^{l}_{M0}\left(\frac{\pi}{2}\right) \times \nonumber\\
&& \sum_{\Mpp=-\lpp}^{\lpp}
d^{\lpp}_{\mpp\Mpp}\left(\frac{\pi}{2}\right) d^{\lpp}_{\Mpp0}\left(\frac{\pi}{2}\right)
\sum_{M^\p=-\lp}^{\lp} d^{\lp}_{mM^\p}\left(\frac{\pi}{2}\right) \, d^{\lp}_{M^\p
m^\p}\left(\frac{\pi}{2}\right)\times \nonumber\\
&& i^{n+m+m'+m''} \, (-1)^{M+M''+M'}
\int_0^\pi \sin\theta d\theta \, e^{i(M+M^\p+\Mpp)\theta}.
\end{eqnarray}

The above expression is real. The proof is given below:
\begin{itemize}
\item Contribution for all of $M,M^\p,\Mpp=0$:

For this term the integral of the above expression is real.
Therefore, if $n+m+m^\prime+\mpp=even$ this term is real (because
then the factor $i^{n+m+m^\prime+\mpp}$ is real). When
$n+m+m^\prime+\mpp=odd$, which means at least one of
$n,m+m^\p,\mpp$ is odd, this term does not contribute, since
$d^l_{m0}(\pi/2) d^l_{0m^\p}(\pi/2)=0$ if $m+m^\p=odd$ (follows
from Eq.~(6) of \S 4.16 of Varshalovich et al. (1988)).

\item Contribution for \emph{not} all of $M,M^\p,\Mpp=0$:

For each set of $M,M^\p,\Mpp$ in the above summation, there
exists a set $-M,-M^\p,-\Mpp$, which converts the integral of the
above expression to its complex conjugate. Since
$d^l_{mm^\p}(\pi/2) =  (-1)^{l-m^\p} d^l_{-mm^\p}(\pi/2)
(-1)^{l+m} d^l_{m-m^\p}(\pi/2)$ (see Eq.~(1) of \S 4.4 of
Varshalovich et al. (1988)), the Wigner-$d$ symbols give a factor
of $(-1)^{n+m+m^\prime+\mpp}$. So, if $n+m+m^\prime+\mpp=even$,
the sum is real, as well as the factor $i^{n+m+m^\prime+\mpp}$ and
both are imaginary if $n+m+m^\prime+\mpp=odd$.
\end{itemize}
Therefore, the full summation is always real.

Following the discussion on the reality of the expression and
using Eq.~(\ref{eq:deff}) we can write
\begin{eqnarray}
J^{l\lpp\lp}_{n\mpp mm^\p} &=&  \delta_{(m-n)\mpp} \, (-1)^m
\frac{\sqrt{(2l+1)(2\lpp+1)}}{2}  \ \times\nonumber\\
&&\sum_{M=-l}^{l} d^{l}_{nM}\left(\frac{\pi}{2}\right)
d^{l}_{M0}\left(\frac{\pi}{2}\right) \sum_{\Mpp=-\lpp}^{\lpp}
d^{\lpp}_{(m-n)\Mpp}\left(\frac{\pi}{2}\right)
d^{\lpp}_{\Mpp0}\left(\frac{\pi}{2}\right)  \ \times\nonumber\\
&&\sum_{M^\p=-\lp}^{\lp}
d^{\lp}_{mM^\p}\left(\frac{\pi}{2}\right) \, d^{\lp}_{M^\p
m^\p}\left(\frac{\pi}{2}\right) \, f_{m'(M+M'+M'')},
\end{eqnarray}
where, as defined in Eq.~(\ref{eq:deff}),
\begin{eqnarray}
f_{m'N} &:=& \Re\left[\Gamma_{m'N}[\rho(\q)=0]\right] \ \equiv \
\Re\left[ i^{m^\p} \, (-1)^N \int_0^\pi
\sin\theta d\theta \, e^{iN\theta} \right] \\
&=& \ \left\{
\begin{array}{cl} (-1)^{(m^\p\pm 1)/2} \, \pi/2        & \mbox{if
$m^\p=odd$ and $N=\pm 1$}\\ (-1)^{m^\p/2}
\, 2/(1-N^2) & \mbox{if both $m^\p,N=0$ or $even$}\\
0               & \mbox{otherwise.} \end{array} \right.
\end{eqnarray}
Note that, for ``symmetric" beams $\beta_{lm} = 0$ for $m=odd$,
so in the final expression terms with $m^\p=odd$ shall not
contribute, that is, \textit{for symmetric beams} $f_{m'N}$
contributes \textit{only} when both $m^\p,N=0$ or $even$.

We can now write the full expression for the bias
matrix for non-rotating beams with incomplete sky coverage in a closed form as:
\begin{eqnarray}
A_{l l^\prime} &=& B_l^2 \, \frac{(2\lp+1)}{16\pi}
\sum_{n=-l}^{l} \sum_{m=-l^\prime}^{l^\prime} \left| \sum_{\lpp
=0}^{\infty} \sqrt{2\lpp+1} \, U_{\lpp(m-n)} \right. \ \times\\
&&\sum_{\Mpp=-\lpp}^{\lpp}
d^{\lpp}_{(m-n)\Mpp}\left(\frac{\pi}{2}\right) \,
d^{\lpp}_{\Mpp0}\left(\frac{\pi}{2}\right) \sum_{M=-l}^{l}
d^{l}_{nM}\left(\frac{\pi}{2}\right) \,
d^{l}_{M0}\left(\frac{\pi}{2}\right) \ \times \nonumber\\
&& \left.  \sum_{M^\p=-\lp}^{\lp}
d^{\lp}_{mM^\p}\left(\frac{\pi}{2}\right)
\sum_{m^\prime=-l^\prime}^{l^\prime} \beta_{l^\prime m^\prime} \,
d^{\lp}_{M^\p m^\p}\left(\frac{\pi}{2}\right) \, f_{m'(M+M'+M'')}
\right|^2.\nonumber
\end{eqnarray}
It is obvious that the above expression is identical to the
expression for equal declination scans
[Eq.~(\ref{eq:BiasWigdEDS})], only the $\Gamma$ coefficients have
been replaced by the $f$ coefficients. If we had started by
substituting $\rho(\theta) = 0$ in Eq.~(\ref{eq:JClebsch}), the
above expression would take the form of
Eq.~(\ref{eq:BiasClebschEDS}) with, again, the $\Gamma$
coefficients replaced by the $f$ coefficients. Which means that,
the computation cost for non-rotating beams and equal declination
scan strategies are the same, only the precomputed $\Gamma$
coefficients for that scan strategy have to be supplied.

\section{Consistency Checks}


\subsection{The full sky and circular beam limit}
\label{cbfs} In this appendix we recover the special case of
circular beam and complete sky coverage limit. From
Eq.~(\ref{eq:BiasClebschEDS}), the full sky limit [$U_{lm} \sqrt{4\pi}\delta_{l0}$] is obtained by replacing $U_{\lpp(m-n)}$
with $\sqrt{4\pi}\delta_{\lpp 0}\delta_{m n}$; and for the
circular beam, we replace the BDP $\beta_{l^\prime m^\prime}$ with
$\delta_{m^\prime 0}$. So, using the definition given in
Eq.~(\ref{eq:deff}),
\begin{eqnarray}
A_{l l^\prime} &=& B_l^2 \, \frac{2\lp+1)}{4}
\sum_{n=-\rm{min}(l,l^\prime)}^{\rm{min}(l,l^\prime)} \left|{
C^{l0}_{l000} \, C^{ln}_{ln00}}  \sum_{\Lp=|l-\lp|}^{l+\lp} C^{\Lp
0}_{l-n\lp n} \, C^{\Lp 0}_{l0\lp 0} \right. \times \nonumber\\ &&
\left.\sum_{N=-\Lp}^{\Lp} d^{\Lp}_{0N}\left(\frac{\pi}{2}\right)
{d^{\Lp}_{N0}\left(\frac{\pi}{2}\right) \, f_{0N}}\right|^2.
\label{allp_semifull}
\end{eqnarray}
From the relation $C^{c\gamma}_{a\alpha
00}=\delta_{ac}\delta_{\alpha\gamma}$ (Eq. (2) in \S 8.5.1 of
\cite{VMK}), we can reduce $C^{l0}_{l000}$ and $C^{ln}_{ln00}$
to unity, and get:
\begin{eqnarray}
A_{l l^\prime} = B_l^2 \, \frac{2\lp+1}{4}
\sum_{n=-\rm{min}(l,l^\prime)}^{\rm{min}(l,l^\prime)}
\left|{\sum_{\Lp=|l-\lp|}^{l+\lp} C^{\Lp 0}_{l-n\lp n} \, C^{\Lp
0}_{l0\lp 0} \sum_{N=-\Lp}^{\Lp}
d^{\Lp}_{0N}\left(\frac{\pi}{2}\right)
{d^{\Lp}_{N0}\left(\frac{\pi}{2}\right) \, f_{0N}}} \right|^2.
\label{allp_semifull-2}
\end{eqnarray}
To get the value of $\sum_{N=-\Lp}^{\Lp}
d^{\Lp}_{0N}\left(\frac{\pi}{2}\right)
{d^{\Lp}_{N0}\left(\frac{\pi}{2}\right) f_{0N}}$, we have to
start a step back.
\begin{eqnarray}
\nonumber
Y_{lm}^*(\hat{q})&=&\sqrt{\frac{2l+1}{4\pi}}D_{m
m^\prime}^l(\hat{q},0)\\
&=&\sqrt{\frac{2l+1}{4\pi}}i^m e^{-im\phi}\sum_{N=-l}^l (-1)^N
d^l_{mN}\left(\frac{\pi}{2}\right)
d^l_{N0}\left(\frac{\pi}{2}\right)e^{iN\theta}
\end{eqnarray}
From the relation
\begin{eqnarray}
\nonumber \int_{4\pi} Y_{lm}^*(\hat{q})d\Omega_{\hat{q}} \sqrt{4\pi} \delta_{l0}\delta_{m0}
\end{eqnarray}
it follows that
\begin{equation}
\sqrt{\frac{2l+1}{4\pi}} i^m \sum_{N=-l}^l (-1)^N
d^l_{mN}\left(\frac{\pi}{2}\right)
d^l_{N0}\left(\frac{\pi}{2}\right) \int e^{iN\theta} \sin\theta
d\theta \int e^{-im\phi}d\phi \ = \ \sqrt{4\pi}
\delta_{l0}\delta_{m0}.
\end{equation}
The last integral $\int e^{-im\phi}d\phi$ gives
$2\pi\delta_{m0}$. So, equating out both sides and rearranging,
we have
\begin{equation}
i^m \sum_{N=-l}^l (-1)^N d^l_{mN}\left(\frac{\pi}{2}\right)
d^l_{N0}\left(\frac{\pi}{2}\right) \int e^{iN\theta} \sin\theta
d\theta \int e^{-im\phi}d\phi = \frac{2}{\sqrt{2l+1}} \delta_{l0}.
\label{dpiby2_sum}
\end{equation}
The l.h.s. is identified with the relation $\sum_{N=-\Lp}^{\Lp}
d^{\Lp}_{0N}\left(\frac{\pi}{2}\right)
{d^{\Lp}_{N0}\left(\frac{\pi}{2}\right) f_{0N}}$, and hence
Eq.(\ref{allp_semifull-2}) reduces to
\begin{eqnarray}
\nonumber A_{l l^\prime} &=& B_l^2 \, \frac{2\lp+1}{4}
\sum_{n=\max(-l,-l^\prime)}^{\min(l,l^\prime)}
\left|{\sum_{\Lp=|l-\lp|}^{l+\lp} C^{\Lp 0}_{l-n\lp n} \, C^{\Lp
0}_{l0\lp 0}
\frac{2}{\sqrt{2L^\prime+1}} \delta_{L^\prime0} }\right|^2 \\
&=& B_l^2 \, (2\lp+1)
\sum_{n=\max(-l,-l^\prime)}^{\min(l,l^\prime)} \left|{ C^{0
0}_{l-n\lp n} \, C^{0 0}_{l0\lp 0} }\right|^2.
\label{allp_semifull-3}
\end{eqnarray}
We know from Eq. (1) of \S 8.5.1 of Varshalovich et al. (1988)
\begin{eqnarray}
\nonumber C^{00}_{a\alpha b\beta}(-1)^{a-\alpha}\frac{\delta_{ab}\delta_{\alpha,
-\beta}}{\sqrt{2a+1}}.
\end{eqnarray}
Hence, $A_{l l^\prime}$ finally reduces to the well known result
\begin{equation}
\nonumber A_{l l^\prime} \ = \ B_l^2 \, \delta_{l \lp}.
\end{equation}


\subsection{The circular beam limit with incomplete sky coverage}
\label{cbcs} We will show in this appendix that our formulation
reduces to the analytic limit of the MASTER method of Hivon et
al. (2002) for the incomplete sky coverage taking circular beams.
Putting $U_{lm} = \sqrt{4\pi}\delta_{l0}$, as done in
appendix~\ref{cbfs}, we get:
\begin{eqnarray}
\nonumber
A_{l l^\prime} &=&
B_l^2 \, \frac{2\lp+1}{16\pi}
\sum_{n=-l}^{l} \sum_{m=-l^\prime}^{l^\prime} \left| {
\sum_{\lpp =0}^{\infty} \sqrt{2\lpp+1} \, U_{\lpp (m-n)}
\sum_{L=|l-\lpp|}^{l+\lpp} C^{L0}_{l0\lpp 0} \,
C^{Lm}_{ln\lpp(m-n)}} \right. \\
&& \left.  \times
\sum_{\Lp=|L-\lp|}^{L+\lp} C^{\Lp 0}_{L-m\lp m} \, C^{\Lp 0}_{L0\lp 0}
\sum_{N=-\Lp}^{\Lp} d^{\Lp}_{0 N}\left(\frac{\pi}{2}\right)
d^{\Lp}_{N 0} \left(\frac{\pi}{2}\right) \, f_{0N}  \right|^2 .
\label{allp_cbcs_1}
\end{eqnarray}
Using Eq. (\ref{dpiby2_sum}), we get
\begin{eqnarray}
\nonumber
A_{l l^\prime}
&=& B_l^2 \,
\frac{2\lp+1}{4\pi}  \sum_{n=-l}^{l} \sum_{m=-l^\prime}^{l^\prime} ~\times \\
\nonumber
&& \left| {
\sum_{\lpp =0}^{\infty} \sqrt{2\lpp+1} \, U_{\lpp (m-n)}
\sum_{L=|l-\lpp|}^{l+\lpp} C^{L0}_{l0\lpp 0} \,
C^{Lm}_{ln\lpp(m-n)}} C^{00}_{L(-m)\lp m} \, C^{00}_{L0\lp 0} \right |^2
\\
\nonumber
&=&
\frac{B_l^2 }{4\pi}
\sum_{n=-l}^{l} \sum_{m=-l^\prime}^{l^\prime} \left| {
\sum_{\lpp =0}^{\infty} \sqrt{2\lpp+1} \, U_{\lpp (m-n)}
\sum_{L=|l-\lpp|}^{l+\lpp} C^{L0}_{l0\lpp 0} \,
C^{Lm}_{ln\lpp(m-n)}} \right|^2
\end{eqnarray}
To arrive at Eq. A31 of Hivon et al. (2002), we first replace
$(m-n)$ by $\mpp$ and then open up the modulus square. The symbol
$C^{l^\prime m}_{ln\lpp \mpp}$ contributes only when $\mpp$ is
equal to $(m-n)$ and also $\lpp$ satisfies the triangle
inequality.
\begin{eqnarray}
\nonumber
A_{l l^\prime} &=& \frac{B_l^2}{4\pi \, (2l^\prime +1) } \sum_{n=-l}^{l}
\sum_{m=-l^\prime}^{l^\prime} \left| { \sum_{\lpp =0}^{\infty}
\sqrt{2\lpp+1} \, \sum_{\mpp =-\lpp}^{\lpp} U_{\lpp \mpp}
C^{l^\prime 0}_{l0\lpp 0} \, C^{l^\prime m}_{ln\lpp \mpp}} \right|^2 \\
\nonumber
&=& \frac{B_l^2}{4\pi(2l^\prime
+1) } \sum_{n=-l}^{l} \sum_{m=-l^\prime}^{l^\prime}\left[
\sum_{l_1^{\prime\prime}=0}^\infty \sqrt{2l_1^{\prime\prime}+1}
\, C^{l^\prime 0}_{l0 l_1^{\prime\prime} 0} \,
\sum_{l_2^{\prime\prime}=0}^\infty \sqrt{2l_2^{\prime\prime}+1} \,
C^{l^\prime 0}_{l0 l_2^{\prime\prime} 0} \,~\times \right. \\
\nonumber
&&  \left. \sum_{m_2^{\prime\prime}
=-l_2^{\prime\prime}}^{l_2^{\prime\prime}}
\sum_{m_1^{\prime\prime}
=-l_1^{\prime\prime}}^{l_1^{\prime\prime}}
U_{l_1^{\prime\prime}m_1^{\prime\prime}} \,
U_{l_2^{\prime\prime}m_2^{\prime\prime}}^* \, C^{l^\prime m}_{ln
l_1^{\prime\prime} m_1^{\prime\prime}} \,C^{l^\prime m}_{ln
l_2^{\prime\prime} m_2^{\prime\prime}} \right] \\
\nonumber
&=& \frac{B_l^2}{4\pi(2l^\prime
+1) } \sum_{l_1^{\prime\prime}=0}^\infty
\sum_{l_2^{\prime\prime}=0}^\infty \, \sqrt{2l_1^{\prime\prime}+1}
\sqrt{2l_2^{\prime\prime}+1} \, C^{l^\prime 0}_{l0
l_1^{\prime\prime} 0} \,
C^{l^\prime 0}_{l0 l_2^{\prime\prime} 0} \,~\times  \\
&&   \sum_{m_2^{\prime\prime}
=-l_2^{\prime\prime}}^{l_2^{\prime\prime}}
\sum_{m_1^{\prime\prime}
=-l_1^{\prime\prime}}^{l_1^{\prime\prime}}
U_{l_1^{\prime\prime}m_1^{\prime\prime}} \,
U_{l_2^{\prime\prime}m_2^{\prime\prime}}^* \, \sum_{n=-l}^{l}
\sum_{m=-l^\prime}^{l^\prime} \, C^{l^\prime m}_{ln
l_1^{\prime\prime} m_1^{\prime\prime}} \,C^{l^\prime m}_{ln
l_2^{\prime\prime} m_2^{\prime\prime}}
\end{eqnarray}
The last summation $\sum_{n=-l}^{l} \sum_{m=-l^\prime}^{l^\prime}
\, C^{l^\prime m}_{ln l_1^{\prime\prime} m_1^{\prime\prime}}
\,C^{l^\prime m}_{ln l_2^{\prime\prime} m_2^{\prime\prime}}$
simplifies to $ (2\lp+1)/(2l_1^{\prime\prime}+1)
\delta_{l_1^{\prime\prime}l_2^{\prime\prime}}
\delta_{m_1^{\prime\prime}m_2^{\prime\prime}}$ by Eq. (5) of
\S~8.7.2 of Varshalovich et al. (1988). So, we have
\begin{eqnarray}
\nonumber
A_{l l^\prime} &=& \frac{B_l^2}{4\pi} \sum_{\lpp=|l-\lp|}^{l+\lp} \,
\left(
C^{l^\prime 0}_{l0 \lpp 0} \right)^2 \, \sum_{\mpp
=-\lpp}^{\lpp} |U_{\lpp\mpp}|^2 \\
&=& B_l^2 \, \frac{2l^\prime +1}{4\pi} \sum_{\lpp=|l-\lp|}^{l+\lp} \,
(2\lpp+1)
\, \wjjj{l}{\lp}{\lpp}{0}{0}{0}^2 \, {\cal{U}}_{\lpp},
\end{eqnarray}
where ${\cal{U}}_{\lpp} \equiv \sum_{\mpp=-\lpp}^{\lpp}
|U_{\lpp\mpp}|^2/(2\lpp+1)$. This matches the final expression of
Hivon et al. (2002) (see Eq. [A31]).


\subsection{The full sky limit with non-circular beam}
\label{ncbfs}

The full sky limit to the final expression should reproduce the
result obtained in Mitra et al. (2004). We substitute $U_{lm} \sqrt{4\pi} \delta_{l0}$ [$\Rightarrow \ U_{l''(m-n)}=\sqrt{4\pi}
\delta_{l''0}\delta_{mn}$] in Eq.~(\ref{eq:BiasClebschEDS}) and
get
\begin{eqnarray}
A_{l l'} &=& B_l^2 \, \frac{(2l'+1)}{4}
\sum_{m=-\rm{min}(l,l')}^{\rm{min}(l,l')} \left| { C^{l0}_{l000}
\, C^{lm}_{lm00}}  \sum_{L=|l-l'|}^{l+l'} C^{L0}_{l-ml'm} \ \times
\right.
\nonumber\\
&& \left.\sum_{N=-L}^{L} d^{L}_{0N}\left(\frac{\pi}{2}\right)
{\sum_{m'=-l'}^{l'} \beta_{l' m'} \, C^{Lm'}_{l0l'm'} \,
d^{L}_{Nm'}\left(\frac{\pi}{2}\right) \, f_{m'N}} \right|^2.
\end{eqnarray}
Using the relation $C^{c\gamma}_{a\alpha00} = \delta_{ac}
\delta{\alpha\gamma}$ (Eq.~(2) in \S 8.5.1 of Varshalovich et al.
(1988)) we may write $C^{l0}_{l000} = C^{lm}_{lm00} = 1$. Then
rearranging terms, we may write
\begin{eqnarray}
A_{l l'} &=& B_l^2 \, \frac{(2l'+1)}{4}
\sum_{m=-\rm{min}(l,l')}^{\rm{min}(l,l')} \left|
\sum_{m'=-l'}^{l'} \beta_{l' m'} \sum_{L=|l-l'|}^{l+l'}
C^{L0}_{l-ml'm} \,
C^{Lm'}_{l0l'm'} \ \times \right. \nonumber\\
&& \left. \sum_{N=-L}^{L} d^{L}_{0N}\left(\frac{\pi}{2}\right) \,
d^{L}_{Nm'}\left(\frac{\pi}{2}\right) \, f_{m'N}\right|^2.
\end{eqnarray}
Using the definition of $f_{m'N}$ [Eq.~(\ref{eq:deff})] and the
expansion formula for Wigner-$d$ [Eq.~(\ref{eq:wigdexp})] we may
write
\begin{equation}
\sum_{N=-L}^{L} d^{L}_{0N}\left(\frac{\pi}{2}\right) \,
d^{L}_{Nm'}\left(\frac{\pi}{2}\right) \, f_{m'N} \ = \ \int_{-1}^{1}
d\cos\theta \, d^L_{0m'}(\theta).
\end{equation}
Then, using the Clebsch-Gordon series [Eq.~(\ref{CGSD})] we get
\begin{equation}
\sum_{L=|l-l'|}^{l+l'} C^{L0}_{l-ml'm} \, d^L_{0m'}(\theta) \,
C^{Lm'}_{l0l'm'} \ = \ (-1)^m \, d^l_{m0}(\theta) \,
d^{l'}_{mm'}(\theta).
\end{equation}
Finally, putting everything together, we get the expression for the bias
matrix in the full sky limit with non-circular beam as
\begin{equation}
A_{l l'} \ = \ B_l^2 \, \frac{(2l'+1)}{4}
\sum_{m=-\rm{min}(l,l')}^{\rm{min}(l,l')} \left|
\sum_{m'=-l'}^{l'} \B_{l' m'} \, \int_{-1}^{1} d\cos\theta \,
d^l_{m0}(\theta) \, d^{l'}_{mm'}(\theta) \right|^2,
\end{equation}
which matches Eq.~(38) of Mitra et al. (2004).


\section{Fast Computation of Bias Matrix for Non-circular Beam in CMB
Analysis}
\label{app:fastCMBBias}

Computation of the bias matrix as given by
Eqs.~(\ref{eq:BiasClebschEDS}) or (\ref{eq:BiasWigdEDS}) in a
naive way is too costly. For fast computation of bias using
Eq.~(\ref{eq:BiasWigdEDS}) we employ a smart algorithm as
described in section~\ref{subsec:factComp}. The details of the
algorithm and cost estimation are given below. Possibility of
fast computation of bias starting from
Eq.~(\ref{eq:BiasClebschEDS}) is being explored.

The full expression for the bias matrix for equal declination
scans, $\rho(\q) \equiv \rho(\theta)$,
[Eq.~(\ref{eq:BiasWigdEDS})]:
\begin{eqnarray}
A_{l l^\prime} &=& B_l^2 \, \frac{2\lp+1}{16\pi}
\sum_{n=-l}^{l} \sum_{m=-l^\prime}^{l^\prime} \left| \sum_{\lpp
=0}^{l_{\rm{mask}}} \sqrt{2\lpp+1} \, U_{\lpp(m-n)} \right. \ \times\\
&&\sum_{\Mpp=-\rm{min}(m_{\rm{mask}},l'')}^{\rm{min}(m_{\rm{mask}},l'')}
d^{\lpp}_{(m-n)\Mpp}\left(\frac{\pi}{2}\right) \,
d^{\lpp}_{\Mpp0}\left(\frac{\pi}{2}\right) \sum_{M=-l}^{l}
d^{l}_{nM}\left(\frac{\pi}{2}\right) \,
d^{l}_{M0}\left(\frac{\pi}{2}\right) \ \times \nonumber\\
&&\left. \sum_{M^\p=-\lp}^{\lp}
d^{\lp}_{mM^\p}\left(\frac{\pi}{2}\right)
\sum_{m^\prime=-\rm{min}(m_{\rm{beam}},l')}^{\rm{min}(m_{\rm{beam}},l')}
\beta_{l^\prime m^\prime} \, d^{\lp}_{M^\p
m^\p}\left(\frac{\pi}{2}\right) \,
\Gamma_{m'(M+M^\p+\Mpp)}[\rho(\theta)] \right|^2. \nonumber
\end{eqnarray}
The following sequence of calculation is computationally cost
effective. $V^{1,2,3}$ have been used as intermediate arrays.
This prescription is only for the loops inside the modulus, so
for each set of $l,l',m,n$ all the three steps have to be
performed.

\begin{itemize}

\item Step I:
\begin{equation}
V_{l\lp}^1[N,m] \ = \ \sum_{M^\p=-\lp}^{\lp}
d^{\lp}_{mM^\p}\left(\frac{\pi}{2}\right)
\sum_{m^\prime=-\rm{min}(m_{\rm{beam}},l')}^{\rm{min}(m_{\rm{beam}},l')}
\beta_{l^\prime m^\prime} \, d^{\lp}_{M^\p
m^\p}\left(\frac{\pi}{2}\right) \,
\Gamma_{m'(M+M^\p+\Mpp)}[\rho(\theta)],
\end{equation}
$N$ runs from $-(l + l_{\rm{mask}})$ to $+(l + l_{\rm{mask}})$.

\item Step II
\begin{equation}
V_{l\lp}^2[\Mpp,n,m] \ = \ \sum_{M=-l}^{l} d^{l}_{nM}\left(\frac{\pi}{2}\right) \,
d^{l}_{M0}\left(\frac{\pi}{2}\right) \, V^1_{l\lp}[M+\Mpp;m].
\end{equation}

\item Step III
\begin{equation}
V_{l\lp}^3[m,n] = \sum_{\lpp =0}^{l_{\rm{mask}}} \sqrt{2\lpp+1} \,
U_{\lpp(m-n)}
\sum_{\Mpp=-\rm{min}(m_{\rm{mask}},l'')}^{\rm{min}(m_{\rm{mask}},l'')}
d^{\lpp}_{(m-n)\Mpp}\left(\frac{\pi}{2}\right) \,
d^{\lpp}_{\Mpp0}\left(\frac{\pi}{2}\right) \, V^2_{ll'}[\Mpp,n,m].
\end{equation}

\end{itemize}
Required number of cycles to compute $V^3$ for each pair $m,n$
(for $l_{\rm{max}} \gg {\lpp}_{\rm{max}}$):
\begin{equation}
\left[\{2(l
+\lpp_{\rm{max}})+1\}(2\lp+1)(2m^{\prime}_{\rm{max}}+1) \, + \,
(2\lpp_{\rm{max}}+1)(2l+1) +
\sum_{\lpp=0}^{\lpp_{\rm{max}}}(2\lpp+1)\right]
\end{equation}
As mentioned earlier we are interested in the total number of
computation cycles in the limit $l_{\rm{max}} \gg l_{\rm{mask}}$.
Before proceeding further we note that $U_{\lpp\mpp} = U_{\lpp
m-n}$ is limited to only $m_{\rm{mask}}$ modes for each $\lpp$.
Here ${m_{\rm{mask}}} > 0$. Then the condition for non zero
$U_{\lpp, m-n}$ becomes $|m-n| < m_{\rm{mask}}$. This in turn
implies that $m-n < m_{\rm{mask}}$ when $ m-n > 0$, and $ -m+n <
m_{\rm{mask}} $ when $ m-n < 0$. Then we see that for each $n, m$
can run only from $n-m_{\rm{mask}}$ to $n+m_{\rm{mask}}$ for a
total of $2m_{\rm{mask}}+1$ values so that $U_{\lpp m-n}$ are non
zero.

Thus considering two outer loops over $m,n$  total computation cycles
becomes
\begin{eqnarray}
&&\sum_{l=2}^{l_{\rm{max}}}\sum_{l'=2}^{l_{\rm{max}}}(2l+1)(2m_{\rm{mask}}+1)\,
\bigg[\{2(l+l_{\rm{mask}})+1\}(2l'+1)(2m'_{\rm{max}}+1) +
\nonumber\\
&& \left. (2l''_{\rm{max}}+1)(2l+1)
+\sum_{l''=0}^{l_{\rm{mask}}}(2l''+1)\right] \nonumber\\
&=& (2m_{\rm{mask}}+1)\sum_{l=2}^{l_{\rm{max}}}
\sum_{l'=2}^{l_{\rm{max}}}(2l+1)\left[\{2(l+l_{\rm{mask}})+1\}
(2l'+1)(2m'_{\rm{max}}+1) + \right. \nonumber\\
&& \left. (2l_{\rm{mask}}+1)(2l+1)
+\sum_{l''=0}^{l_{\rm{mask}}}(2l''+1)\right]\nonumber\\
&=& (2m_{\rm{mask}}+1)\sum_{l=2}^{l_{\rm{max}}}
\sum_{l'=2}^{l_{\rm{max}}}(2l+1)\left[2l_{\rm{mask}}(2l'+1)(2m'_{
\rm{max}}+1)+ \right. \nonumber\\
&& \left. (2l +1)(2l'+1)(2m'_{\rm{max}}+1) +
(2l''_{\rm{max}}+1)(2l+1)
+\sum_{l''=0}^{l''_{\rm{max}}}(2l''+1)\right]
\end{eqnarray}
The computation cycles will be decided by the maximum power of the
largest term in the above expression. Clearly the second term in the
bracket will give the maximum contribution as it contains highest powers
combined from $l,l'$. Hence the total number of cycles is
\begin{eqnarray}
&&(2m_{\rm{mask}}+1)\sum_{l=2}^{l_{\rm{max}}}4l^2\sum_{l'=2}^{l_{\rm{max}}}(2l')(2m'_{\rm{max}}+1)
\ = \
(2m_{\rm{mask}}+1)(2m'_{\rm{max}}+1) \ \times\nonumber\\
&& \ \ \
8\left[\frac{l_{\rm{max}}(l_{\rm{max}}+1)(2l_{\rm{max}}+1)}{6}
-1\right]\left[\frac{l_{\rm{max}}(l_{\rm{max}}+1)}{2}-1\right]
\end{eqnarray}
For $l_{\rm{max}} \gg 1$  the computation cost scales as $(4/3)
(2m_{\rm{mask}}+1)(2m'_{\rm{max}}+1)l^5_{\rm{max}} $.

\section{Expansion of Wigner-D Function}

\subsection{Motivation}

This derivation is motivated from Eq.~(10) of \S 4.16 of
Varshalovich et al. (1988). However, the motivating equation had
certain inconsistency, as it predicts
$D^l_{mm^\prime}(\phi,\theta,\rho) = 0$ if $m+m^\prime$ is $odd$,
which, in general, is not true. We rectify the formula by
``reverse engineering''. We start with the second expression of
the above mentioned equation [see below for steps]:
\begin{eqnarray}
&&\sum_{M_1,M_2,M_3,M_4=-l}^{l} \left[ D^l_{mM_1}\left(\phi,0,0\right)
\, D^l_{M_1M_2}\left(0,\frac{\pi}{2},0\right) \,
D^l_{M_2M_3}\left(0,\theta,0\right) ~\times \right.\nonumber\\
&&\left. \ \ \ D^l_{M_3M_4}\left(0,\frac{\pi}{2},0\right) \,
D^l_{M_4m^\prime}\left(0,0,\rho\right) \right] \nonumber\\
&=& e^{-im\phi} \sum_{M_2,M_3=-l}^{l} \left[
D^l_{mM_2}\left(0,\frac{\pi}{2},0\right) \,
D^l_{M_2M_3}\left(\theta,0,0\right) \,
D^l_{M_3m^\prime}\left(0,\frac{\pi}{2},0\right) \right]
e^{-im^\prime\rho} ~~~\rm{[Step ~1]} \nonumber\\
&=& e^{-im\phi} \sum_{M_2=-l}^{l} \left[
D^l_{mM_2}\left(0,\frac{\pi}{2},0\right) \,
D^l_{M_2m^\prime}\left(\theta,\frac{\pi}{2},0\right) \right]
e^{-im^\prime\rho} ~~~~~~~~~~~~~~~~~~~~~~~~~~~~\rm{[Step ~2]}
\nonumber\\
&=& e^{-im\phi} \,
D^l_{mm^\prime}\left(\frac{\pi}{2},\pi-\theta,\frac{\pi}{2}\right)
\, e^{-im^\prime\rho}
~~~~~~~~~~~~~~~~~~~~~~~~~~~~~~~~~~~~~~~~~~~~~~~~~~~~~~~\rm{[Step ~3]}
\nonumber \\
&=& D^l_{mm^\prime}\left(\frac{\pi}{2}+\phi,\pi-\theta,\frac{\pi}{2}
+\rho\right)
\label{eq:motive}
\end{eqnarray}

\subsection{Details of the Steps in the above derivation}

\begin{itemize}
\item{Step I}:

From Eq.~(1) \& (2)  of \S 4.16, pg.112 of Varshalovich et al.
(1988).
\begin{eqnarray}
D^l_{mm^\prime}(\phi,0,0) &=& e^{-im\phi} \, D^l_{mm^\prime}(0,0,0) \\
D^l_{mm^\prime}(0,0,\rho) &=& D^l_{mm^\prime}(0,0,0) \,
e^{-im^\prime\rho}\\
D^l_{mm^\prime}(0,0,0) &=& \delta_{mm^\prime}.
\end{eqnarray}

\item{Step II}:

From the ``Addition of Rotations" formula in Eq.~(3) of \S 4.7,
pg.87 of Varshalovich et al. (1988).
\begin{equation}
\sum_{M=-l}^{l} \left[ D^l_{mM}(\phi,\theta_1,\gamma) \,
D^l_{Mm^\prime}(-\gamma,\theta_2,\rho) \right] \ = \
D^l_{mm^\prime}(\phi,\theta_1+\theta_2,\rho).
\end{equation}

Another way is to combine the first two remaining $D$ symbols
using Eq.~(1) of \S 4.16, pg.112 of Varshalovich et al. (1988)
and then evaluate the following in Step~III using the ``Addition
of Rotations" formula similar to the present method:
\begin{equation}
e^{-im\phi} \sum_{M_3=-l}^{l} \left[
D^l_{mM_3}\left(0,\frac{\pi}{2},\theta\right) \,
D^l_{M_3m^\prime}\left(0,\frac{\pi}{2},0\right) \right]
e^{-im^\prime\rho}.
\end{equation}

\item{Step III}:

From Eq.~(1) of \S 4.7, pg.87 of Varshalovich et al. (1988) we
may write
\begin{equation}
\sum_{M=-l}^{l} \left[ D^l_{mM}(0,\frac{\pi}{2},0) \,
D^l_{Mm^\prime}(\theta,\frac{\pi}{2},0) \right] \ = \
D^l_{mm^\prime}(\alpha,\beta,\gamma)
\end{equation}
where $\alpha, \beta, \gamma$ are to be obtained using
Eq.~(66)-(70) of \S 1.4, pg.32 of Varshalovich et al. (1988).
Note that the arguments of the \emph{first} $D$ symbol have been
denoted by $\alpha_2, \beta_2, \gamma_2$ respectively and
\emph{not} by
$\alpha_1, \beta_1, \gamma_1$.\\
From Eq.~(66) of \S 1.4, pg.32 of Varshalovich et al. (1988),
since $0\le\alpha< 2\pi, 0\le \beta\le \pi, 0\le\gamma< 2\pi$
\begin{eqnarray}
\cos\alpha \ = \ 0 &\Rightarrow& \alpha \ = \  \frac{\pi}{2}\ \mbox{or}
\ \frac{3\pi}{2}\\
\cos\beta \ = \ -\cos\theta &\Rightarrow& \beta \ = \ \pi \, - \,
\theta\\
\cos\gamma \ = \ 0 &\Rightarrow& \gamma \ = \ \frac{\pi}{2} \
\mbox{or} \ \frac{3\pi}{2}.
\end{eqnarray}
From Eq.~(67) of \S 1.4, pg.32 of Varshalovich et al. (1988)
\begin{equation}
\sin\alpha \ = \ \sin\gamma \ = \ \frac{\sin\theta}{\sin\theta} \
= \ 1.
\end{equation}
Combining the above equations we may write
\begin{equation}
\alpha \ = \ \frac{\pi}{2}; \ \ \ \beta \ = \ \pi \, - \, \theta;
\ \ \ \gamma \ = \ \frac{\pi}{2}.
\end{equation}
\end{itemize}

\subsection{Final expression}

We can modify Eq.~(\ref{eq:motive}) by changing $\phi \rightarrow
\phi-\frac{\pi}{2}, \theta \rightarrow \pi-\theta, \rho
\rightarrow \rho-\frac{\pi}{2}$ to reach the desired expansion:
\begin{eqnarray}
&&D^l_{mm^\prime}\left(\phi,\theta,\rho\right) \ = \
e^{-im(\phi-\pi/2)} e^{-im^\prime(\rho-\pi/2)} \ \times\\
&&\sum_{M_2,M_3=-l}^{l} \left[
D^l_{mM_2}\left(0,\frac{\pi}{2},0\right) \,
D^l_{M_2M_3}\left(\pi-\theta,0,0\right) \,
D^l_{M_3m^\prime}\left(0,\frac{\pi}{2},0\right) \right]. \nonumber
\end{eqnarray}
Then using the definitions of Wigner-$d$ functions from Eq.~(1) of
\S 4.3, pg.76 and Eq.~(1) of \S 4.16, pg.112 of Varshalovich et
al. (1988), we get
\begin{equation}\
D^l_{mm^\prime}\left(\phi,\theta,\rho\right) \ = \ i^{m+m^\prime}
\, e^{-im\phi} \sum_{M=-l}^{l} \left[ (-1)^M \,
d^l_{mM}\left(\frac{\pi}{2}\right) \, e^{iM\theta} \,
d^l_{Mm^\prime}\left(\frac{\pi}{2}\right) \right]
e^{-im^\prime\rho}.
\end{equation}
This also means
\begin{equation}
d^l_{mm^\prime}(\theta) \ = \ i^{m+m^\prime} \sum_{M=-l}^{l}
\left[ (-1)^M \, d^l_{mM}\left(\frac{\pi}{2}\right) \,
e^{iM\theta} \, d^l_{Mm^\prime}\left(\frac{\pi}{2}\right)
\right].\label{eq:wigdexp}
\end{equation}
The coefficients $d^l_{mm^\prime}(\pi/2)$ can be directly
calculated using Eq.~(5) of \S 4.16, pg.113 of Varshalovich et
al. (1988)
\begin{eqnarray}
d^l_{mm^\prime}\left(\frac{\pi}{2}\right) &=& (-1)^{m-m^\prime}
\, \frac{1}{2^l}  \,
\sqrt{\frac{(l+m)!(l-m)!}{(l+m^\prime)!(l-m^\prime)!}} \ \times\\
&&\sum_{k=\max\{0,\, m^\prime-m\}}^{\max\{l+m^\prime,\, l-m\}}
(-1)^k \, \mat{l+m^\prime\\k} \, \mat{l-m^\prime\\k + m -
m^\prime}.\nonumber
\end{eqnarray}


\section{Useful formulae}
\label{appendix:uf}

\begin{itemize}

\item Important relations [Eq.~(1)s of \S 4.3, \S 4.17 \& \S 5.4
and Eq.~(2) of \S 4.4 of Varshalovich et al. (1988)]
\begin{eqnarray}
D_{mm^\prime}^l(\mathbf{\hat{q}},\rho) &=& e^{-im\phi} \,
d_{mm^\prime}^l(\theta) \, e^{-im^\p\rho} \label{Dd} \\
Y^*_{lm}(\mathbf{\hat{q}}) &=& \sqrt{\frac{2l+1}{4\pi}} \,
D_{m0}^{l}(\mathbf{\hat{q}},\rho) \, = \, \sqrt{\frac{2l+1}{4\pi}} \,
e^{-im\phi} \, d_{m0}^l(\theta) \label{YD} \\
D_{mm^\prime}^{l*}(\mathbf{\hat{q}},\rho) &=& (-1)^{m-m^\p} \,
D_{-m-m^\prime}^{l}(\mathbf{\hat{q}},\rho) \label{DstarD}\\
Y^*_{lm}(\mathbf{\hat{q}}) &=& (-1)^m \,
Y_{l-m}(\mathbf{\hat{q}}) \label{YstarY}
\end{eqnarray}
Note that, unlike Mitra et al. (2004), the argument of the
Wigner-$d$ function is $\theta$ (standard definition) \emph{not}
$\cos\theta$.

\item The Clebsch-Gordon series:

Expansion of the product of two Wigner-$D$ functions [Eq~(1) of
\S 4.6 of Varshalovich et al. (1988)]:
\begin{eqnarray}
&&D^{l_1}_{m_1n_1}(\q,\rho) \, D^{l_2}_{m_2n_2}(\q,\rho) \ = \nonumber\\
&& \ \ \ \sum_{l=|l_1-l_2|}^{l_1+l_2} C^{l(m_1+m_2)}_{l_1m_1l_2m_2} \,
D^{l}_{(m_1+m_2)(n_1+n_2)}(\q,\rho) \,
C^{l(n_1+n_2)}_{l_1n_1l_2n_2}, \label{CGSD}
\end{eqnarray}
where $C^{lm}_{l_1m_1l_2m_2}$ are the Clebsch-Gordon coefficients.\\
The special case of spherical harmonics [Eq.~(9) of \S 5.6 of
Varshalovich et al. (1988)]:
\begin{eqnarray}
&&Y_{l_1m_1}(\q) \, Y_{l_2m_2}(\q) \ = \label{CGSY}\\
&&\ \ \ \sum_{l=|l_1-l_2|}^{l_1+l_2}
\sqrt{\frac{(2l_1+1)(2l_2+1)}{4\pi(2l+1)}} \, C^{l0}_{l_10l_20}
\, C^{l(m_1+m_2)}_{l_1m_1l_2m_2} \,
Y_{l(m_1+m_2)}(\q).\nonumber
\end{eqnarray}
In modifying the above equations (from Varshalovich et al.
(1988)) we have used the fact that the Clebsch-Gordon
coefficients $C^{lm}_{l_1m_1l_2m_2}$ vanish if $m \ne m_1+m_2$.

\end{itemize}



\end{document}